# The semi-classical stress-energy tensor in a Schwarzschild background, the information paradox, and the fate of an evaporating black hole

James M. Bardeen

*Physics Department, Box 1560, University of Washington*
*Seattle, Washington 98195-1560. USA*
*bardeen@uw.edu*

**Abstract**

Analytic approximations to and numerical results for the semi-classical stress-energy tensor outside the horizon of a Schwarzschild black hole obtained in the 1980's and 1990's are re-examined in order to better understand the origin of Hawking radiation and the implications for the black hole information paradox. Polynomial fits to the numerical results for the tangential stress in 4D are obtained for conformally coupled spin 0 and spin 1 fields in the Hartle-Hawking and Unruh quantum states. What these results show is that the origin of the Hawking radiation is not pair creation or tunneling very close to the black hole horizon, but rather is a nonlocal process extending beyond the potential barrier in the mode propagation equations centered around $r = 3M$. Arguments are presented that the black hole information paradox cannot plausibly be addressed by processes occurring close to the horizon of a black hole whose geometry is close to Schwarzschild. However, a toy model for the evolution of the black hole geometry based on the Bousso covariant entropy bound suggests a possible resolution.

## I. INTRODUCTION

The original derivation of Hawking radiation from black holes[1] and the prediction of its essentially thermal character at the Hawking temperature

$$T_\text{H} = \frac{\kappa m_\text{p}^2}{2\pi} \quad \left(G = c = 1, \hbar = m_\text{p}^2\right), \tag{1.1}$$

was based on semi-classical effective field theory. The surface gravity of the horizon $\kappa$ is $1/(4M)$ for a Schwarzschild black hole of mass $M$. I do not set the Planck mass $m_\text{p}$ equal to one, in order to emphasize the smallness of quantum corrections for a large astrophysical black hole, of order $m_\text{p}^2 / M^2 < 10^{-76}$ for $M > 1 M_\odot$. In the semi-classical approximation the quantum fields propagate in a classical background spacetime, a solution of the classical Einstein equations. The Hawking luminosity of a massless field of spin $s$ has been parameterized as

$$L_\text{H} = \frac{4\pi}{245760\pi^2} \frac{m_\text{p}^2}{M^2} k_s = 4\pi M^2 \sigma T_\text{H}^4 k_s, \tag{1.2}$$

where $\sigma = \pi^2 / (60 m_p^6)$ is the Stefan-Boltzmann constant. For a solar mass or larger black hole only spin 1 photons and spin 2 gravitons are expected to contribute, for which Page[2] has calculated $k_1 = 6.4928$ and $k_2 = 0.7404$, respectively. A hypothetical conformally coupled massless scalar field has $k_0 = 14.36$ (Elster[3]; Taylor, Chambers, and Hiscock[4]). Later calculations determined the expectation value of the complete renormalized semi-classical stress-energy tensor (SCSET) for conformally coupled spin 0 and spin 1 fields outside the horizon, $r > 2M$, using the point-splitting renormalization procedure of Christensen[5] and applied to black holes by Christensen and Fulling (CF)[6]. The standard assumptions are that the SCSET is time-independent (neglecting the back-reaction on the geometry) and satisfies local energy-momentum conservation. The trace of the SCSET is zero classically, but the renormalization breaks the conformal invariance of the fields to produce an anomalous trace depending only on the local curvature of the background spacetime and the spin of the field.

    The calculations require specifying the quantum state. While historically this has been called a choice of vacuum, this is an unfortunate misnomer, since what is "vacuum" is ill-defined in an inhomogeneous curved spacetime for modes with wavelengths the order of the curvature scale. An unambiguous definition of particles as excitations of a "vacuum" is only possible in the approximately Minkowskian spacetime well before formation of the black hole or in the asymptotically flat region far from the black hole. The two states usually considered are the Hartle-Hawking (HH) state[7], the thermal state for an eternal black hole in equilibrium with an external heat bath at the Hawking temperature, and the Unruh state[8] appropriate for a black hole formed by gravitational collapse. which evolves from the initial approximately Minkowski vacuum. The Hartle-Hawking state is static, with no net energy flux as seen by a static observer anywhere outside the future and past horizons. The Unruh state has no incoming radiation at past null infinity and outgoing radiation at future null infinity. Both are regular in a freely falling frame at the future horizon. The Unruh state SCSET is calculated assuming transients associated with formation of the black hole have decayed, so the SCSET is approximately stationary on time scales small compared with the evaporation time. The quasi-stationary approximation is well justified, since for a large black hole the evaporation time is enormously longer than (of order $M^2 / m_p^2$ times) the dynamical relaxation time scale of order $M$. A third state sometimes considered is the Boulware state, which is static like the HH state, but with a SCSET falling off faster than $r^{-2}$ at infinity and singular at $r = 2M$. This is the appropriate state for the exterior of a spherical star in static equilibrium, which must have a radius $R > 2M$, not a black hole.

    Numerical results for the tangential stress $\langle T_\theta^\theta \rangle$ of a conformally coupled spin 0 field in the HH state in a Schwarzschild background were obtained by Howard and Candelas[9] (HC), and later with improved accuracy by Anderson, Hiscock and Samuel (AHS)[10]. Elster[11] attempted an extension to the spin 1 HH state, but made an error in his application of point-splitting renormalization. Corrected spin 1 HH results were obtained by Jensen and Ottewill (JO)[12]. Elster's calculation[3]

of the difference between the Unruh and HH SCSETs for a spin 0 field was more successful. Jensen, McLaughlin, and Ottewill[13] (JMO) improved on his results and extended them to the spin 1 case. It is only necessary to calculate the tangential stress component $<T_\theta^\theta>$ from scratch, since the conservation laws and the trace anomaly determine the remaining components of the SCSET. Unfortunately, while tabular data for the spin 0 HH and Unruh states was preserved in a paper by Visser[14], the spin 1 numerical results were published only as rather crude graphs.

Analytic expressions fitted to the numerical results facilitate their physical interpretation. Visser[14] found a fit to the tangential stress of a conformally coupled scalar field in the Unruh vacuum as a polynomial in $2M/r$. However, his fit, taking into account the constraints implied by the conservation of the SCSET, is not quite consistent with the value of the spin 0 Hawking luminosity quoted above. In Part II of the paper I consider carefully the framework for such fits and using the data from Visser obtain remarkably precise polynomial fits to the tangential stress for the spin 0 Hawking and Unruh states. I also extract numerical values and construct polynomial fits from the published graphs for the spin 1 Hawking and Unruh states and the spin 0 Boulware state. There are limits to how accurately this can be done, so the latter fits are more ambiguous. All of my fits are obtained using the Levenberg-Marquardt nonlinear least squares algorithm.

The fits to the numerical results are compared with analytic *approximations* in the literature, which were pioneered by Page[15]. He made use of the fact that an 'ultrastatic" spacetime obtained by conformal transformation from a static Ricci-flat geometry like Schwarzschild has no trace anomaly. Page argued that this justified a simple approximation to the HH SCSET in the ultrastatic geometry, which could then be conformally transformed back to the physical spacetime. Brown and Ottewill[16] derived the Page approximation from an effective action split into two parts, an "anomalous" part whose variation with respect to the metric defines an anomaly stress-energy tensor (ASET), and a "thermal" part invariant under conformal transformations. The anomalous trace in the physical spacetime, the trace of the ASET, vanishes in the ultrastatic spacetime, and it was assumed that the full ASET vanishes there as well. Both the ASET and the thermal contribution are singular on the horizon in the physical spacetime, but the net SCSET is regular and reproduces the Page approximation. While the Page approximation worked rather well as originally derived for the spin 0 HH state, an attempt by Brown, Ottewill, and Page (BOP)[17] to extend the approximation to the spin 1 HH state failed badly. JO improved the BOP spin 1 HH approximation by adding an extra contribution not derived from an action. Matyjasek suggested approximations to the spin 0[18] and spin 1[19] Unruh state SCSETs based in part on the JO HH approximation and in part on numerical fits to the data.

The physical interpretation of the SCSET is taken up in Part III, with the primary focus on the Unruh state, as appropriate for a "real" black hole. The Unruh state energy flux in static frames is at large $r$ an outward flow of positive energy, the Hawking radiation, and at the horizon is an inward flow of negative energy. How and where the transition between these limits occurs is an indication of where the Hawking radiation is being generated. I argue that the SCSET is inconsistent with the Hawking radiation being generated very close to the horizon. The conversion of vacuum fluctuations into Hawking radiation is something that happens nonlocally and at least for a scalar field is centered around $r = 3M$, not the horizon at $r = 2M$.

In Part IV, I discuss the first-order back-reaction on the geometry associated with the SCSET. The accurate results for the SCSET confirm the qualitative picture of black hole evaporation widely accepted in the literature, and the conclusion of Bardeen[20] that the metric remains Schwarzschild to a very good approximation on and outside the horizon as the black hole evaporates. What happens in the interior of the black hole is more speculative.

Part V considers of the implications for the black hole information "paradox". I emphasize that that the notion of an *event* horizon trapping information forever is a classical concept, relying on energy conditions that are violated by the SCSET. I comment on some attempts (e.g., by Bousso, et al[21]) to provide the basis for *quantum* singularity theorems and find them not compelling. Taking into account quantum backreaction on the geometry, I see no reason to think that a quantum black hole must have an event horizon. Therefore, whenever I refer simply to the "horizon" I just mean an apparent horizon.

Storing a significant amount of quantum information in a non-degenerate "stretched horizon"[22] or in a "thermal atmosphere" is not possible as long as the backreaction is small and the semi-classical approximation is valid. The stretched horizon is just a way of dealing with *external* perturbations of a classical black hole. The "thermal atmosphere" is not a property of the black hole, it is just a property of an accelerating particle detector, whether near a black hole horizon or in ordinary Minkowski spacetime. Trapped quantum information ends up in the deep interior of a Schwarzschild black hole on a dynamical time scale, and is not available to resolve the information paradox, assuming it propagates causally in a Schwarzschild background geometry.

On the assumption that quantum backreaction keeps the interior geometry of the black hole nonsingular, and inspired by the Bousso covariant bound[23], I suggest a scenario in which entanglement between the trapped quantum information in the interior and the exterior Hawking radiation drives an enhanced quantum backreaction on the geometry, beyond semi-classical expectations. An inner apparent horizon expands to a radius comparable to the radius of the outer apparent horizon by the Page time. An explicit toy model for the evolution of the metric, elaborated and improved from a previous version[24], is presented in Part VI.

## II. THE SEMI-CLASSICAL STRESS-ENERGY TENSOR OUTSIDE THE SCHWARZSCHILD HORIZON

The SCSET is the expectation value of the renormalized energy-momentum tensor of quantum fields calculated to first-order in $\hbar$ on a fixed classical background geometry, taken here to be the spherically symmetric Schwarzschild geometry, with the static (for $r > 2M$) metric

$$ds^2 = -(1 - 2M/r)dt^2 + (1 - 2M/r)^{-1} dr^2 + r^2(d\theta^2 + \sin^2\theta \, d\varphi^2). \qquad (2.1)$$

At $r > 2M$ it is convenient to work with the physical components of the SCSET as projected onto the orthonormal frames of static observers, uniformly accelerating observers whose world lines are at constant Schwarzschild radius r. However, it is important to remember that static observers are unphysical in the limit $r \to 2M$, where their proper acceleration is infinite. The only physical frames at $r = 2M$ are "falling frames". The global geometry is not static.

The only quantum states considered here are those for which the SCSET is spherically symmetric, with four independent components, an energy density $E = -T_t^t$, an energy flux/momentum density $F = -T_t^r/(1 - 2M/r)$, a radial stress $P_r = T_r^r$, and a tangential stress $P_t = T_\theta^\theta = T_\varphi^\varphi$, all as defined in the frame of a static observer. Outside the horizon any classical disturbances not protected by global conservation laws dissipate by a combination of radiation out to future null infinity or inward across the horizon to the black hole interior, unless generated by external sources. After transient behavior associated with black hole formation, on a time scale of order several times $M$, the expectation value of the energy-momentum tensor of a quantum field is assumed to become stationary to first-order in $\hbar$. As an expectation value, the SCSET should be considered an average over times very long compared with $M$, but very short compared with the evaporation time of order $M^3/m_\text{p}^2$. The SCSET must be conserved, since it is constructed to be a source in the classical Einstein equations. With time derivatives set to zero, the energy conservation equation is

$$r^2 T_{t;\beta}^\beta = \partial_r \left[ r^2 \left(1 - \frac{2M}{r}\right) F \right] = 0 \qquad (2.2)$$

and momentum conservation is

$$r T_{r;\beta}^\beta = (E + P_r)\frac{M/r}{1 - 2M/r} + \frac{1}{r}\partial_r(r^2 P_r) - 2P_t = 0. \qquad (2.3)$$

Since the Hawking luminosity $L_\text{H} = \lim_{r \to \infty}(4\pi r^2 F)$, we see from Eqs. (1.2) and (2.2) that the spin $s$ contribution to the net energy flux at finite $r$ is

$$F = \frac{k_s}{245760} \frac{m_\text{p}^2}{M^2 r^2} \frac{1}{(1 - 2M/r)} = k_s \sigma T_\text{H}^4 \frac{M^2}{r^2} \frac{1}{(1 - 2M/r)}. \qquad (2.4)$$

For classically conformally invariant quantum fields, the only contribution to the trace of the SCSET is the trace (i.e. Weyl) anomaly. The renormalization of the quantum fields breaks the classical conformal invariance, leading to a non-zero trace depending only on the curvature tensors of the classical background geometry, and independent of the quantum state[25]. In the Schwarzschild background, the Ricci tensor $R_{\alpha\beta}$ and the scalar curvature $R$ vanish, and for spin $s$

$$T_\alpha^\alpha = q_s \frac{m_\text{p}^2}{2880\pi^2} \left( R_{\alpha\beta\gamma\delta} R^{\alpha\beta\gamma\delta} = 48\frac{M^2}{r^6} \right). \qquad (2.5)$$

The coefficient is $q_0 = 1$ for a conformally coupled massless spin 0 field, $q_{1/2} = 7/2$ for a massless spin 1/2 (Dirac) field, $q_1 = -13$ for the spin 1 electromagnetic field, and $q_2 = 212$ for a massless spin 2 field. Then, with $x \equiv 2M/r$,

$$T_\alpha^\alpha = -E + P_r + 2P_t = 64 q_s \sigma T_H^4 x^6. \tag{2.6}$$

My approach is to focus solely on the physical spacetime, in which the numerical calculations are done, and apply the constraints imposed by the conservation of the SCSET. Momentum conservation relates the radial stress and the transverse stress, so the calculation of the full SCSET reduces to calculating one degree of freedom, usually taken to be the transverse stress.

From Eq. (2.4) the static frame energy flux, if nonzero anywhere, is infinite at the horizon. This is perfectly consistent with a SCSET that is non-singular in a falling frame, since the static frame is moving outward at the speed of light relative to any frame freely falling from a nonzero distance outside the horizon. Any ingoing energy in the free-fall frame is infinitely blueshifted in the Lorentz transformation to the static frame, just an any outgoing energy is infinitely redshifted. The infinite boost can produce infinite energy density and radial stress, but leaves the tangential stress unaffected. In order to isolate the singular behavior on the horizon, define an "ingoing" part of the SCSET, $\langle T_\beta^\alpha \rangle^{\text{in}}$, which in the static frame is defined by

$$E^{\text{in}} = P_r^{\text{in}} = -F^{\text{in}} = -F = -\frac{1}{4} k_s \sigma T_H^4 \frac{x^2}{1-x}, \quad P_t^{\text{in}} = 0. \tag{2.7}$$

This, by itself, satisfies momentum conservation, as can be verified from Eq. (2.3), and describes an ingoing radially propagating negative energy null fluid. It vanishes for the HH state.

I call the remainder of the SCSET the "regular" part, and for either the HH state or the Unruh state approximate its radial and transverse stress by the polynomials

$$P_r^{\text{reg}} = h_s P_0 \sum_{n=0}^{N} r_n x^n, \quad P_t^{\text{reg}} = h_s P_0 \sum_{n=0}^{N} t_n x^n. \tag{2.8}$$

The quantity $P_0 \equiv 2\sigma T_H^4 / 3$ is the thermal pressure per helicity state for a massless field at the Hawking temperature, and $h_s$ is the number of helicity states for spin $s$, $h_s = 1$ for a spin 0 field and $h_s = 2$ for a spin 1 field. The regular part is by definition conserved and has zero energy flux in the static frame. The regular energy density is then $E^{\text{reg}} = P_r^{\text{reg}} + 2P_t^{\text{reg}} - T_\alpha^\alpha$. Momentum conservation for the regular part gives relations between the $t_n$ and $r_n$, which are also subject to physically appropriate asymptotic conditions as $r \to \infty$, i.e. $x \to 0$, that depend on the quantum state. The remaining degrees of freedom are determined by fitting to the numerical data. The order of the polynomial is chosen to be the minimum required for a good numerical fit, at least $N = 6$ in order to accommodate the trace anomaly. In these units the trace anomaly is $T_\alpha^\alpha = 96 q_s x^6 P_0$, or $96 x^6 P_0$ for spin 0 and $-1248 x^6 P_0$ for spin 1.

The result of substituting Eqs. (2.8) into the momentum conservation equation is

$$(n-2)r_n = -2t_n + (n-2)r_{n-1} + 3t_{n-1} - (48q_s/h_s)\delta_{n7}. \tag{2.9}$$

Through $n = 8$:

$$\begin{aligned}
&r_0 = t_0, \; r_1 = 2t_1 - 2t_0, \; t_2 = \frac{3}{2}t_1, \; r_3 = r_2 + \frac{9}{2}t_1 - 2t_3, \\
&r_4 = r_2 + \frac{9}{2}t_1 - \frac{1}{2}t_3 - t_4, \; r_5 = r_2 + \frac{9}{2}t_1 - \frac{1}{2}t_3 - \frac{2}{3}t_5, \\
&r_6 = r_2 + \frac{9}{2}t_1 - \frac{1}{2}t_3 + \frac{1}{12}t_5 - \frac{1}{2}t_6, \; r_7 = r_2 + \frac{9}{2}t_1 - \frac{1}{2}t_3 + \frac{1}{12}t_5 + \frac{1}{10}t_6 - \frac{48q_s}{5h_s} - \frac{2}{5}t_7, \\
&r_8 = r_2 + \frac{9}{2}t_1 - \frac{1}{2}t_3 + \frac{1}{12}t_5 + \frac{1}{10}t_6 + \frac{1}{10}t_7 - \frac{48q_s}{5h_s} - \frac{1}{3}t_8.
\end{aligned} \tag{2.10}$$

The condition that the polynomials terminate at $n = N$ is $r_{N+1} = t_{N+1} = 0$, which for $N = 6$ implies that

$$t_6 = -10r_2 - 45t_1 + 5t_3 - \frac{5}{6}t_5 + \frac{96q_s}{h_s}, \tag{2.11}$$

for $N = 7$

$$t_7 = -10r_2 - 45t_1 + 5t_3 - \frac{5}{6}t_5 - t_6 + \frac{96q_s}{h_s}, \tag{2.12}$$

and for $N = 8$

$$t_8 = -\frac{21}{2}\left(r_2 + \frac{9}{2}t_1 - \frac{1}{2}t_3 + \frac{1}{12}t_5 + \frac{1}{10}t_6 + \frac{1}{10}t_7 - \frac{48q_s}{5h_s}\right). \tag{2.13}$$

Together these equations guarantee that the regularity condition $\left(E^{\text{reg}} + P_r^{\text{reg}}\right)_{x=1} = 0$ is satisfied identically. Of course, a polynomial representation implicitly assumes a smooth finite limit as $x \to 1$, as is confirmed by the numerical calculations for both the HH and Unruh states.

The free parameters in Eqs. (2.10) are $r_2$ and the $t_n$ for $n \neq 2$ and $n < N$. The asymptotic conditions on the SCSET appropriate in the HH state, for which $\langle T_\alpha^\beta \rangle^{\text{in}} \equiv 0$, correspond to equilibrium of a thermal gas in the static frame,

$$P_r = P_t = h_s P_0 (1-x)^2 = h_x P_0 \left(1 + 2x + 3x^2 + \ldots\right). \tag{2.14}$$

This requires $t_0 = 1$, $t_1 = 2$, and $r_2 = 3$. Deviations from the thermal gas expansion are expected at order $x^3$, due to geodesic deviation of the particle trajectories caused by the background curvature. There was some confusion about this point in the early literature, in which there seemed to be an expectation that the first deviations should go as the square of the curvature, at order $x^6$, apparently inspired by the form of the Page analytic approximation for the spin 0 HH state.

For the Unruh state the asymptotic SCSET is dominated by the radial outflow of the Hawking radiation, since there is no incoming radiation. Therefore, $t_0 = t_1 = 0$ and

$$E \cong P_r \cong F \to \frac{1}{4}k_s\sigma T_H^4 x^2 = \frac{3}{8}k_s P_0 x^2. \tag{2.15}$$

This requires $P_r^{\text{reg}} \cong 2F$, and thus

$$r_2 = \frac{3k_s}{4h_s}, \tag{2.16}$$

jn order to compensate for $P_r^{\text{in}} = -F$. While it would seem plausible that the first nonzero $t_n$ should be $t_4$, as one would expect for classical radiation with a small angular spread about the radial direction, the numerical results suggest that $t_3 \neq 0$.

Numerical calculations of the tangential stress have typically proceeded by combining the result for the HH state, $<T_\theta^\theta>_H$ with a calculation of the difference $<T_\theta^\theta>_U - <T_\theta^\theta>_H$ between the Unruh state and the HH state to obtain $<T_\theta^\theta>_U$. The difference calculation is much simpler than calculating the Unruh values directly. The published graphs of the differences are a more sensitive test of the differences in the polynomial coefficients than graphs of the SCSET components for each state separately, particularly for non-zero spins, but with the caveat that at smaller values of $x$ it becomes increasingly difficult to extract accurate Unruh state values from the difference.

### a) Spin 0 Hartle-Hawking state

The Page analytic approximation[15] to the total spin 0 HH tangential stress, including the trace anomaly, is

$$P_t = P_0\left(1 + 2x + 3x^2 + 4x^3 + 5x^4 + 6x^5 - 9x^6\right). \tag{2.17}$$

While often cited as remarkably accurate, this differs from the numerical results of Howard and Candelas[9] and of Anderson, et al[10] by about 50% around $x \sim 0.8$, and by about 17% at $x = 1$. The differences are small compared with the scale set by the trace anomaly. The approximation could easily be improved to agree with the Candelas result for $P_t(1)$ by setting $t_4 = 3.27$. A change in $t_4$ does not require any compensating change in the other $t_n$.

The numerical results of AHS have been preserved in Table II of Visser[14] for $2 < r/M \leq 5$, making it possible to construct an accurate polynomial fit to the numerical data. Matyjasek[18] made an attempt based on the "strong thermal hypothesis", that, as in the Page approximation, there are no corrections to the thermal expansion until $O(x^6)$. This requires an $N = 8$ polynomial to get a reasonable fit, with essentially one free parameter after applying conservation constraints and fixing $P_t(1) = 10.27 P_0$, as derived by Candelas[26]. The result is

$$P_t = P_0\left(1 + 2x + 3x^2 + 4x^3 + 5x^4 + 6x^5 - 83.064x^6 + 108.664x^7 - 36.33x^8\right). \tag{2.18}$$

The maximum error is about 0.7%, with $\chi^2 = 7.04 \times 10^{-3}$.

Allowing a departure from the thermal gas expansion at order $x^3$ results in a remarkably better $N = 6$ fit

$$P_t = P_0\left(1+2x+3x^2+3.650x^3+14.398x^4-48.170x^5+34.392x^6\right) \quad (2.19)$$

with two free parameters, maximum errors of 0.03%, and $\chi^2 = 1.35\times 10^{-5}$. The formal uncertainty in the fitted value of $t_3$ is only 0.005. That such a good fit is possible with two free parameters is testimony both to the accuracy of the AHS numerical results and to how closely this fit must match the exact result. Allowing one more free parameter, in a $N = 7$ fit, reduces the $\chi^2$ only slightly, to $\chi^2 = 1.23\times 10^{-5}$, and the uncertainty in the coefficients becomes several times larger, with a value of $t_7$ consistent with zero. Applying Eqs. (2.10) to Eq. (2.19),

$$P_r = P_0\left(1+2x+3x^2+4.70x^3-4.223x^4+42.288x^5-11.035x^6\right) \quad (2.20)$$

and

$$E = P_0\left(3+6x+9x^2+12x^3+24.573x^4-54.052x^5-38.251x^6\right). \quad (2.21)$$

The comparison of this fit with the numerical data in Fig. 1 does not do justice to its accuracy.

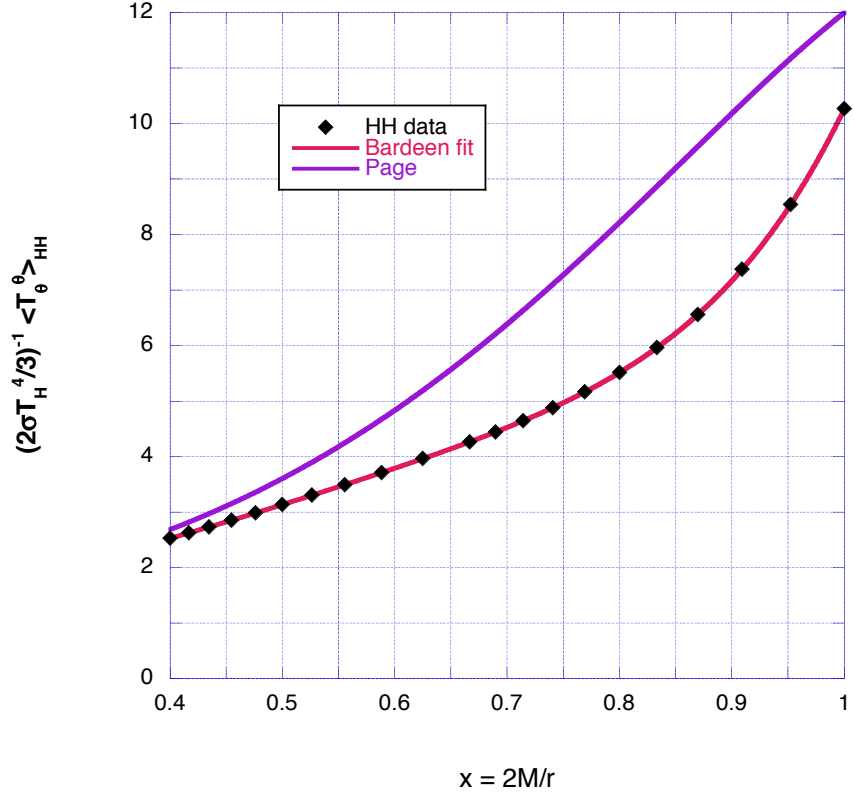

Figure 1. Comparing the numerical data for the spin 0 HH transverse stress with my polynomial fit and the Page analytic approximation.

b) Spin 0 Unruh state

The polynomial fit found by Visser[14] to the spin 0 Unruh state tangential stress. based on the AHS numerical results for the HH state and the JMO results for the Unruh - HH difference, with the lowest power $x^4$ as argued by CF[6], is

$$P_t = P_0 \left( 26.562 x^4 - 59.0214 x^5 + 38.0268 x^6 \right). \tag{2.22}$$

The maximum residual is about 0.7% around $x = 0.5$. This fit made no use of the Hawking luminosity, so $r_2$ as well as $t_5$ and $P_t(x=1)$ were determined just from the tangential stress data. However, the calculation of the Hawking luminosity is simpler and should be more accurate. The spin 0 luminosity found originally by Elster[3] is $L_H = 7.44 \times 10^{-5} \left( m_p / M \right)^2$, which implies $k_0 = 14.36$ and $r_2 = 10.77$. This value was confirmed to the full three significant figures by independent calculations of Simkins[27] and of Taylor, Chambers and Hiscock[28]. The Visser fit implies $k_0 = 14.26$. While the difference seems small, it is significant, because the $10 r_2$ term directly tied to the Hawking luminosity and the other terms in Eq. (2.11) for $t_6$ nearly cancel, making the net result quite sensitive to errors in $r_2$.

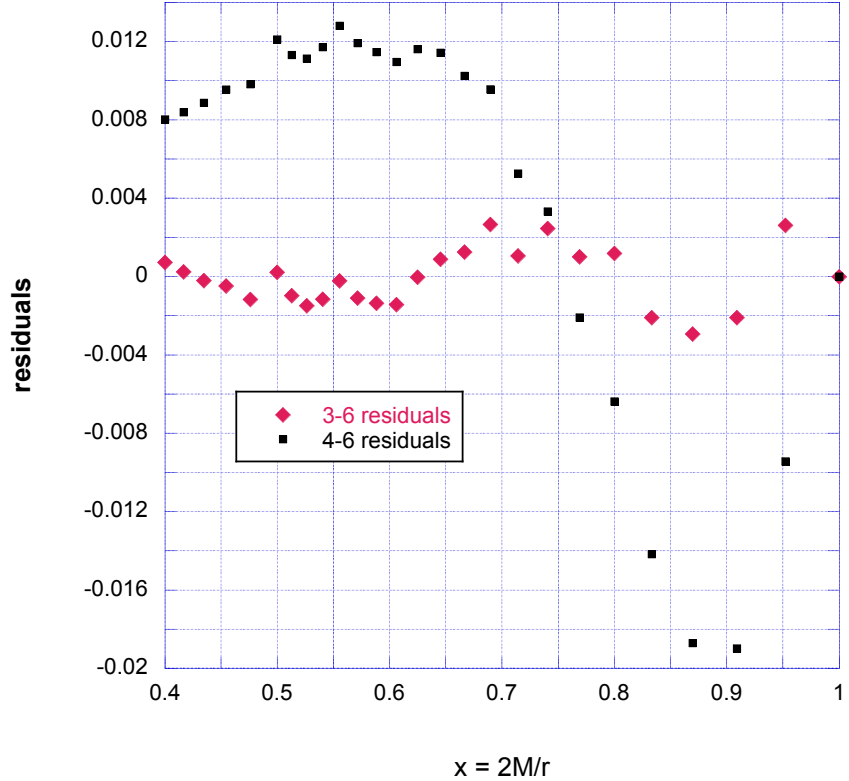

Figure. 2. Comparison of residuals from fitting the data given in Table 2 of Visser for the spin 0 Unruh state tangential stress to a $n = 3-6$ polynomial and a $n = 4-6$ polynomial, consistent with the Elster $k_0 = 14.36$, in each case optimizing $P_t(1)$.

Adopting $r_2 = 10.77$ and fitting with $P_t(1)$ and $t_5$ as the free parameters gives $t_4 = 27.879$, $t_5 - 62.304$, $t_6 = 40.220$, $P_t(1) = 5.795 P_\infty$, with a mediocre $\chi^2 = 2.09 \times 10^{-3}$. Allowing a nonzero $t_3$ results in a much better fit,

$$P_t = P_0 \left(0.264 x^3 + 25.438 x^4 - 57.460 x^5 + 37.503 x^6 \right), \qquad (2.23),$$

with $P_t(1) = 5.745 P_0$ and $\chi^2 = 5.53 \times 10^{-5}$. The formal uncertainty in $t_3$ is $\pm 0.008$ and in $t_5$ is $\pm 0.13$. The residuals of this Unruh state fit are not as small as in the 3-6 HH state fit, presumably because of loss of precision in extracting the Unruh state values at smaller $x$ from the dominant HH state values. In each case $P_t(1)$, not part of the Visser data set, is chosen so the residual at $x = 1$ vanishes. The residuals of the two fits are compared in Figure 2. More accurate numerical calculations extending to larger radii would definitively settle the issue. Allowing $r_2$ to float in the 3-6 fit gives $r_2 = 10.87 \pm 0.29$ and does not substantially reduce the $\chi^2$.

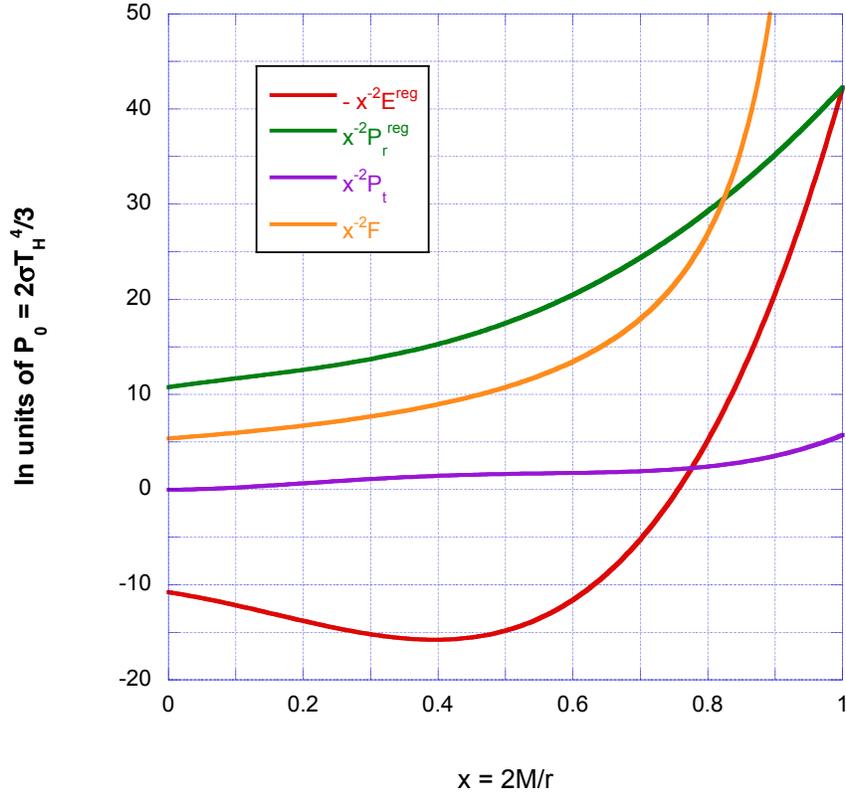

Figure 3. The static frame components of the spin 0 Unruh state SCSET, with the parts of $E$ and $P_r$ associated with the singular energy flux at the horizon removed, and scaled to show the approach to outgoing Hawking radiation as $x \to 0$.

From Eq. (2.23), the "regular" part of the radial stress is

$$P_r^{\text{reg}} = P_0 \left(10.77 x^2 + 10.242 x^3 - 14.800 x^4 + 48.945 x^5 - 12.902 x^6 \right), \qquad (2.24)$$

and the regular part of the energy density is
$$E^{\text{reg}} = P_0\left(10.77x^2 + 10.77x^3 + 36.076x^4 - 65.975x^5 - 33.896x^6\right). \qquad (2.25)$$
All of the components of the SCSET for spin 0 Unruh state are shown together in Fig. 3. Within the range of the numerical data, $x \geq 0.4$, the errors from the 3-6 fit are much less than the width of the lines.

The Matyjasek approximation to spin 0 Unruh state $P_t$ is an $n = 4 - 6$ polynomial[18]. The values of $t_5$ and $t_6$ were assumed to be identical to those of the Page approximation for the HH state, with $t_5 = 6$ and $t_6 = -9$. Momentum conservation then requires $r_2 = 10$, which is inconsistent with the Hawking luminosity. His $t_4 = 8.75$ is chosen to make $P_t(1) = 5.75 P_\infty$, as found numerically. The Maryjasek $t_n$ are very different from those of Eq. (2.23) and Eq. (2.22). Overall, the Matyjasek Unruh approximation is similar in accuracy to the Page HH approximation.

### d) Spin 1 Hartle-Hawking state

JO[12] found a reasonably good analytic approximation to the HH $P_t$, similar in accuracy to the spin 0 Page approximation, but based only in part on effective action arguments. Then JMO[13] calculated the difference of the Unruh and HH tangential stresses. In both cases the results were only presented as graphs, rather than numerical tables. Unfortunately, the original numerical data are no longer available.

The JO analytic approximation to the spin 1 HH state tangential stress is an $N = 6$ polynomial,
$$P_t^{\text{HA}} = 2P_0\left(1 + 2x + 3x^2 + 44x^3 - 305x^4 + 66x^5 - 579x^6\right). \qquad (2.26)$$
The numerical result for the difference between this analytic approximation and the numerical tangential stress, $-\Delta_\theta^\theta$, is plotted in Fig. 4 of JO, with a vertical range from 0 to $720 P_0$ in order to accommodate a comparison with earlier very poor attempts at analytic approximations by Zel'nikov and Frolov[29] and by BOP[17]. The numerical results only range from 0 to about $24 P_0$. The JO $\Delta_\theta^\theta$ is indistinguishable from zero once $r/M > 4.5$ or so, and at best the numerical uncertainties are of order 5%. They did confirm that $\Delta_\theta^\theta$ is tends toward zero at the horizon. The JO analytic approximation already departs quite strongly from the strong thermal hypothesis at the $x^3$ term of the polynomial.

There is no hope of getting an unique polynomial fit to $\Delta_\theta^\theta$ given the form of the available data. The best that can be done is to choose the simplest possible fit that has a reasonably small $\chi^2$ and reasonably well-defined coefficients. A $N = 6$ fit over the range $0.5 \leq x \leq 1$ that does about as well as can be expected, while constraining the coefficients to be consistent with momentum conservation, is
$$\Delta_\theta^\theta = 2P_0\left(-6.89x^3 + 106.20x^4 - 389.17x^5 + 289.86\right). \qquad (2.27)$$
with $\chi^2 = 0.24$. The fit of Eq. (2.27) is compared with the numerical data in Fig. 4. Adding this to Eq. (2.26) gives for the spin 1 HH tangential stress

$$P_t = 2P_0\left(1 + 2x + 3x^2 + 37.11x^3 - 198.80x^4 - 323.17x^5 - 289.14x^6\right). \quad (2.28)$$

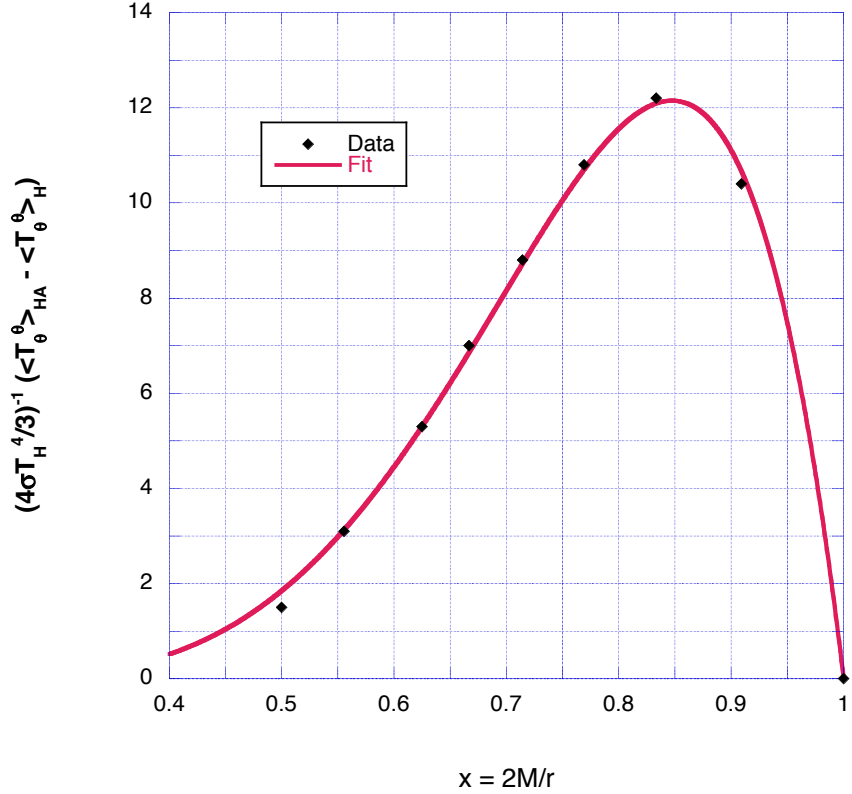

Figure 4. The diamonds are data points for $-\Delta_\theta^\theta$ extracted from Fig. 4 of JO, and the curve is the fit given in Eq. (2.27). Uncertainties in extracting the data from the JO graph are becoming quite large for $x \leq 0.5$.

### d) Spin 1 Unruh state

I get the best results for my polynomial fit to the spin 1 Unruh state tangential stress by combining the data for $\Delta_\theta^\theta$ and the Unruh-HH difference, extrapolating $\Delta_\theta^\theta$ a bit (where it is very small) so the fit can extend over the full $0.4 \leq x \leq 1$ range of the U - HH data, and then fitting $P_t^U - P_t^{HA}$ to a polynomial. The result of a 3-7 fit is

$$P_t = 2P_0\left(57.31x^3 - 593.12x^4 + 696.98x^5 - 1268.71x^6 + 326.10x^7\right), \quad (2.29)$$

with $\chi^2 = 0.234$. The formal uncertainties in the coefficients are rather large, 10% to 30%, so this result should only be taken as a convenient representation of the numerical data, and is not necessarily close to any exact analytic result. To make the differences of the Unruh expressions and the accuracies of the fits more apparent, I plot the data, my fit, and the Matyjasek model for $P_t^U - P_t^{HA}$ in Fig. 5. The differences between the models and accuracy of the fit are much less apparent if the HH analytic approximation is not subtracted. This is because largely state-independent vacuum polarization effects (including the trace anomaly) dominate near the horizon for spin 1.

Using Eqs. (2.10) and (2.12), together with $r_2 = 2.435$ and Eq. (2.29), gives

$$P_r^{reg} = P_0 \left(4.87x^2 - 224.37x^3 + 1133.8x^4 - 981.74x^5 + 1332.42x^6 - 326.1x^7\right) \quad (2.30)$$

and

$$E^{reg} = P_0 \left(4.87x^2 + 4.87x^3 - 1238.68x^4 + 1806.18x^5 - 2494.4x^6 + 978.28x^7\right). \quad (2.31)$$

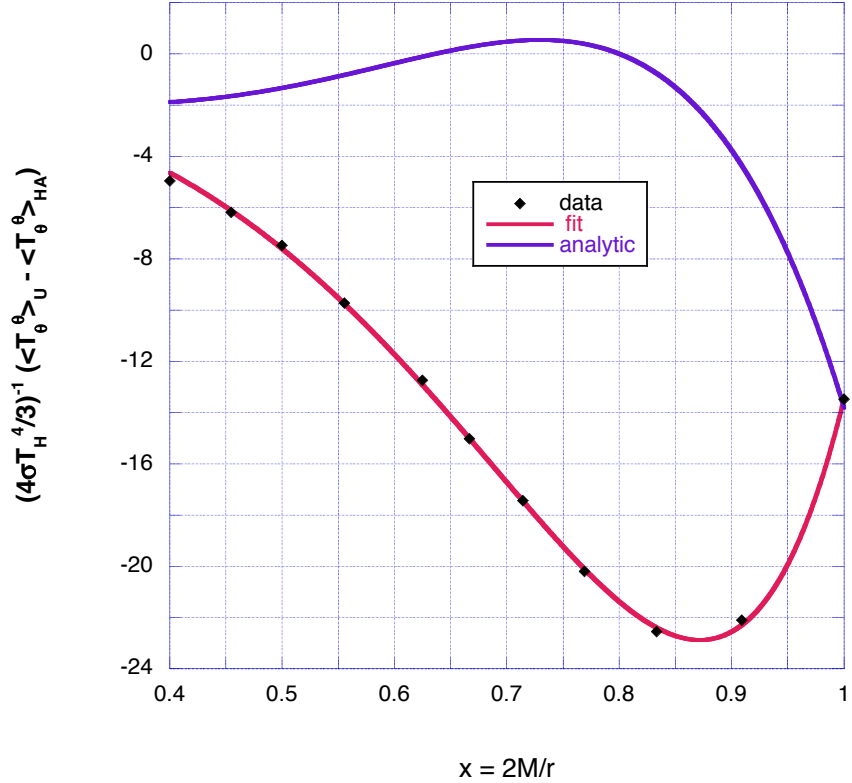

Figure 5. The data points for the spin 1 Unruh state tangential stress extracted from the JO and JMO graphs are plotted after subtracting the JO analytic approximation for the HH state, along with the fit of Eq. (2.29) and the "analytic" Unruh state model of Matyjasek[19].

All of the static frame components are plotted in Fig. 6. What is glaringly obvious, in contrast to spin 0, is how insignificant the energy flux is in comparison with the diagonal components of the SCSET. This is because vacuum polarization effects, as indicated by the trace anomaly, increase by a factor of 13 in magnitude, while the Hawking luminosity decreases by more than a factor of two. Also note the difference of sign for $P_t$ and the enormous difference in magnitude from the spin 0 case. The signs of $E^{reg}$ and $P_r^{reg}$ at the horizon are the same for the two spins, in spite of the difference in sign of the trace anomaly.

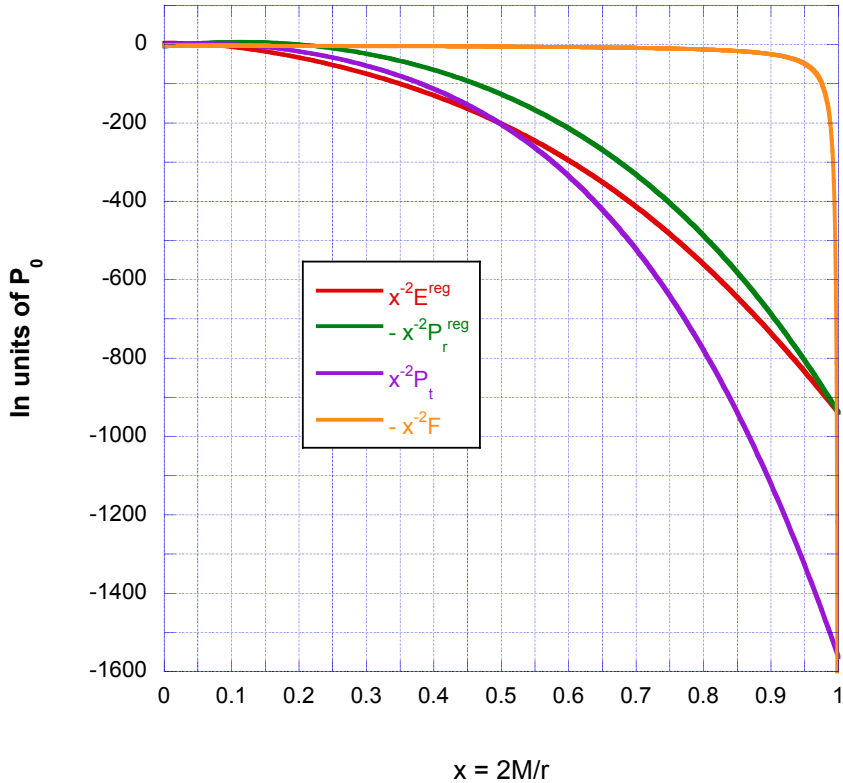

Figure 6. The components of the spin 1 SCSET for the Unruh state. Note that the signs on $E^{\text{reg}}$ and $P_r^{\text{reg}}$ are opposite from Fig. 3.

In the spin 1 fits the values of the coefficients are not nearly as tightly constrained as the deviation of the fitting function from the numerical values. Because there are large cancellations between terms in the polynomials in the region just outside the horizon, varying one coefficient can be compensated by correlated variations of the others with relatively little affect on the net value of the function for $x<1$.

Extrapolation of the polynomials to the interior of the black hole, $x>1$, is not at all justified. While the expectation is that the SCSET should vary smoothly across the horizon for the states being considered, there is no expectation of analyticity and no expectation of an accurate extrapolation of the polynomial fits. For a large quasi-stationary black hole the apparent horizon is very close to being a null hypersurface, and is actually timelike for an evaporating black hole, so data from outside the horizon does not uniquely determine the solution inside.

### e) Work in progress

Levi and Ori[30] have recently embarked on a program to greatly improve the accuracy and range of calculations of SCSETs for black holes. They extend the point-splitting renormalization technique from the separations in $t$ of earlier calculations to separations in the $\theta$ and $\varphi$ directions, in order to accommodate axisymmetric and evolving background geometries. Their initial results were for a *minimally* (not conformally) coupled scalar field in the Schwarzschild background. They confirm an asymptotic $x^3$ falloff of the tangential stress by calculating out to $r = 40M$ all components of the SCSET. They also confirm the ingoing flow of negative energy at the horizon. Their techniques should eventually make it possible to calculate the SCSET during the collapse to form the black hole.

### III. Physical interpretation of the SCSET

A positive energy flux, as in the Unruh state SCSET or the MV ASET, can be due to either positive energy flowing out or negative energy flowing in. At large $r$ only positive energy outflow is physically acceptable. At the horizon only negative energy inflow in the static frame is consistent with a regular SCSET in a falling frame. If I had assumed outflow of positive energy just outside the horizon, as would have been appropriate if Hawking radiation were generated by pair creation within a Planck distance or so of the horizon, and therefore had defined the "regular" part of the SCSET by subtracting off an outgoing radial flow of positive energy, the coefficient $r_2$ in $P_r^{\mathrm{reg}}$ would be zero, since then the outflow of positive energy at infinity would be completely assigned to the "singular" part of the SCSET. However, as we noted in Part II, this is inconsistent with the numerical results for the spin 0 tangential stress, which require $r_2 = 10.8 \pm 0.3$.

To get a better understanding of the transition from inflow at the horizon to outflow at large $r$, define an "outgoing" part of the net energy flux by

$$F^{\mathrm{out}} \equiv \frac{1}{4}(E + P_r + 2F) = \frac{1}{4}(E^{\mathrm{reg}} + P_r^{\mathrm{reg}}) = \frac{1}{2}(P_r^{\mathrm{reg}} + P_t^{\mathrm{reg}}) - \frac{1}{4}T_\alpha^\alpha \equiv \frac{1}{2}(1-x)Z_s. \quad (3.1)$$

The "ingoing" part of the energy flux is

$$F^{\mathrm{in}} \equiv F - F^{\mathrm{out}} = \frac{1}{4}(-E - P_r + 2F). \quad (3.2)$$

This strictly makes sense only if the SCSET is made up of radially propagating null fluids plus a radial boost invariant "vacuum polarization" contribution, but the ratio $F^{\mathrm{out}}/F$ is a useful diagnostic as long as the energy flux is a major part of the SCSET. The quantity $Z_s$ varies smoothly across the horizon and is helpful in making the Lorentz transformation from the static frame to a freely falling frame (see below). Evaluating $Z_s(x)$ using the Unruh state polynomial fits of Part II gives for spin 0

$$Z_0 = P_0\left(10.77x^2 + 21.276x^3 + 31.914x^4 + 23.399x^5\right), \quad (3.3)$$

and for spin 1

$$Z_1 = P_0\left(4.87x^2 - 104.88x^3 - 157.32x^4 + 254.90x^5 - 326.10x^6\right). \quad (3.4)$$

The "outgoing" fraction of the net flux in the static frame is

$$F^{\mathrm{out}}/F = (1-x)^2 Z_s / (h_s r_2 x^2). \quad (3.5)$$

In the Matyjasek approximations[18,19]

$$Z_0^{\mathrm{M}} = P_0\left(10x^2 + 20x^3 + 30x^4 + 42x^5\right), \quad (3.6)$$

$$Z_1^M = P_0\left(4.87x^2 + 9.74x^3 + 14.61x^4 + 36x^5\right). \qquad (3.7)$$

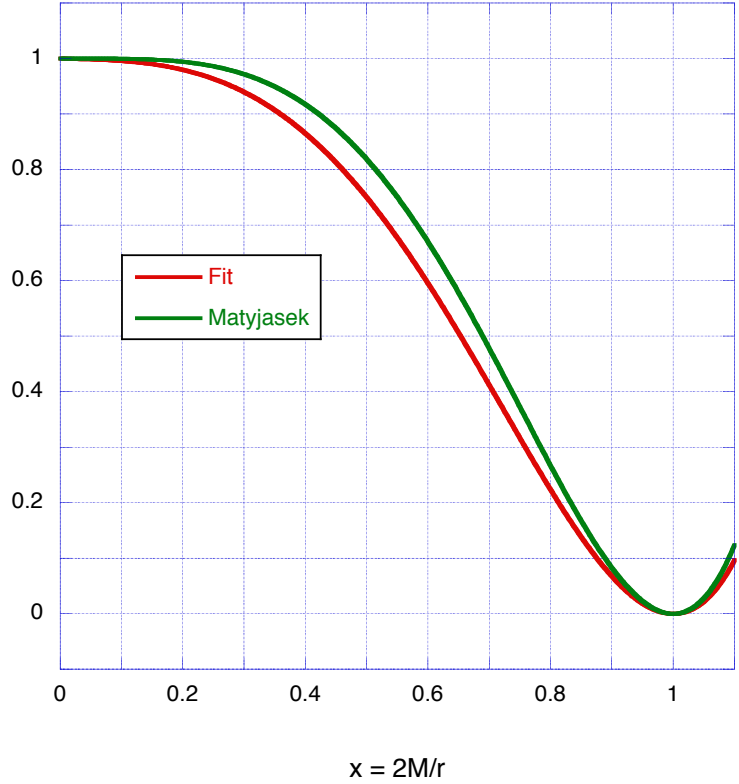

x = 2M/r

Figure 7. The outgoing fraction of the energy flux in the static frame for the spin 0 Unruh state, comparing my fit from the numerical data and the Matyjasek analytic model.

The energy flow interpretation of the spin 0 outgoing fraction for my fit to the numerical data and the analytic approximations seems to work pretty well, as shown in Fig. 7, with positive energy outgoing Hawking radiation dominating the net energy flux where $F^{out}/F > 1/2$, which for spin 0 is for $r > 3M$, $x < 2/3$, outside the peak of the potential barrier in the mode equations. Inside the potential barrier the energy flow is dominated by ingoing negative energy. This is not consistent with a physical picture in which the Hawking radiation, as has often been suggested in the literature, is due to pair creation or tunneling very close to the horizon.

The situation is considerably more complicated for a spin 1 field, as shown in Fig. 8. The "outgoing fraction" is substantially less than zero over much of the vicinity of the black hole, which in the radial energy flow interpretation implies an outflow of negative energy as well as an inflow. However, the Hawking energy flux is so small compared with the rest of the SCSET for spin 1 that the radial energy flow interpretation is not justified. Vacuum polarization effects dominate $E^{reg} + P_r^{reg}$ out to very large radii, with the added caveat that the fit may not reliable where it is an extrapolation of the numerical data, i.e., for $x < 0.5$.

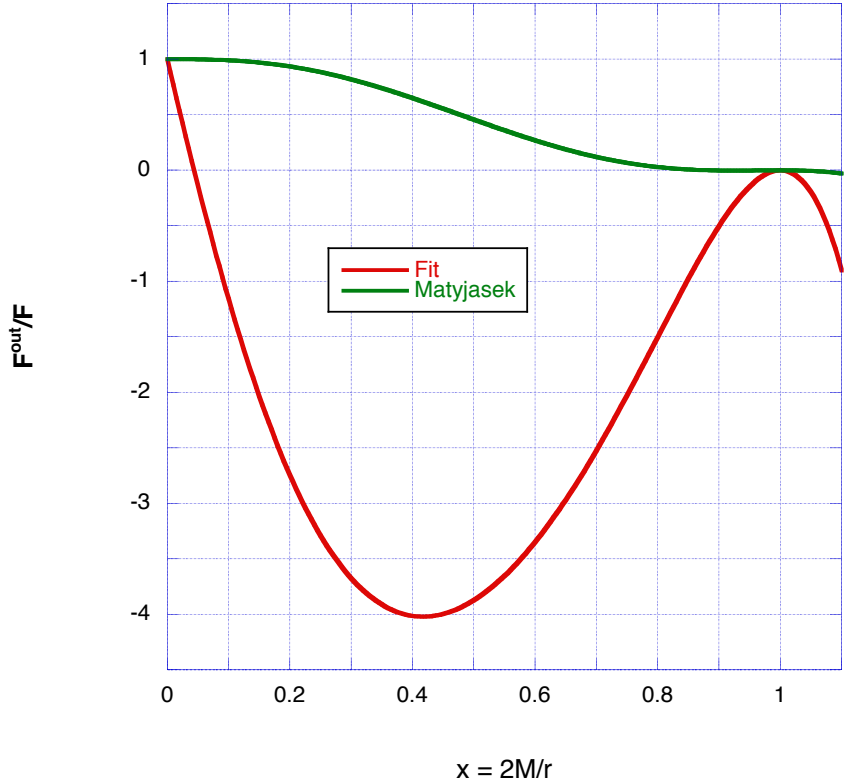

Figure 8.  The nominal outgoing fraction of the spin 1 Unruh state energy flux.  The result from my fit to the numerical data is compared with that from the Matyjasek analytic approximation.

What the spin 0 SCSET implies is that as vacuum fluctuations propagate outward from the horizon, and are partially transmitted through and partially reflected from the potential barrier around $r = 3M$, they should be interpreted as physical particles only at larger radii, ultimately with respect to the Minkowski vacuum at future null infinity.  The ingoing flow of negative energy across the horizon should not be considered as associated with real "particles" or Hawking "partners".  Well inside the horizon it may or may not be appropriate to interpret the vacuum fluctuations as "Hawking partner" particles with locally positive energy and negative Killing energy.  This is the interpretation of the generation of Hawking radiation originally proposed in papers by Unruh[31] and by Fulling[32].

There are good physical reasons why pair creation or tunneling creating Hawking radiation extremely close to the horizon (as proposed, for instance, by Parikh and Wilczek[33]) doesn't make sense, besides being incompatible with the SCSET.  A Hawking particle just outside the horizon must have a large energy and momentum in the static frame and an enormously larger energy in a typical local free-fall frame in order to reach infinity even with the small energy corresponding to the Hawking temperature.  The large local energy is *not* compensated by gravitational potential energy, since this has no local significance by the equivalence principle.  The Hawking partner, in a local pair creation process, as a real particle, must have a corresponding large energy and outward momentum in a local inertial frame straddling the horizon, in order for the partner to have negative Killing energy. This requires an enormous violation of local conservation of energy and momentum.

The existence of Hawking radiation is due to the positive frequency vacuum modes at $\Im^-$ evolving into a mixture of positive and negative frequency modes vacuum modes at $\Im^+$, as argued eloquently by Hawking[1], but this says little about exactly where in the general vicinity of the black hole the particles are created, since the definition of particles is highly ambiguous when their wavelengths are comparable to the curvature scale, as are the wavelengths of modes near the peak of the Hawking spectrum.

It is convenient to calculate the mismatch between the in and out vacuums by extrapolating the mode functions to the horizon of a stationary black hole. Visser[34] has given a nice pedagogical discussion based on Gullstrand-Painleve coordinates $(\tilde{t}, r)$ in Schwarzschild, which are regular on the horizon. In a WKB approximation, the phase of an outgoing scalar mode of frequency $\omega$ is $\mp i \left( \omega \tilde{t} - \int^r k(r') dr' \right)$. Near the horizon $(r - r_H \ll \kappa^{-1})$ the wave number $k \approx \omega / [\kappa (r - r_H) + i\varepsilon]$, with $\kappa$ the surface gravity of the horizon. The wave number diverges and changes sign at the horizon, since outgoing modes become ingoing in radius inside the horizon. The Feynman $i\varepsilon$ ensures the proper phase relation between a Hawking mode just outside the horizon and the "partner" mode just inside. However, any physical excitation is a wave packet, integrated over a range of frequencies. The rapid oscillation of phase near the horizon means that any such wave packet will have very small amplitude close to the horizon due to destructive interference between neighboring frequencies.

The short wavelength modes of a quasi-Minkowski vacuum state remain unexcited as long as they are localized very close to the horizon, where they are just propagating in the locally flat geometry. However, they are increasing redshifted relative to the uniformly accelerating static observers of the Schwarzschild spacetime, as they would be relative to Rindler observers in a flat spacetime. As their wavelengths become comparable to the radius of the black hole geodesic deviation effects become large. Far from the black hole it becomes possible to interpret the outgoing modes as physical particles relative to a different local vacuum. Modes contributing to the high frequency tail of the Planck spectrum at the Hawking temperature never experience very much geodesic deviation, and contribute very little to the asymptotic Hawking energy flux.

Is there any basis for the recent claim of Baker, et al[35] and others that there is significant entanglement of Hawking radiation with quantum fluctuations of the background geometry? Their argument is based on the huge energy of Hawking quanta created infinitesimally close to the horizon, but if Hawking quanta are created around $r = 3M$ with energy $\sim m_p^2 / M$ spread out over a distance of order $M$ the time delay induced by the backreaction as a fraction of the period of the Hawking wave is of order $\left(m_{\text{p}}^2 / M\right) M / M^2 \sim m_{\text{p}}^2 / M^2 \sim 10^{-76}$ for a solar mass black hole. The modes that give rise to Hawking radiation well after the formation of the black holes do have sub-Planckian wavelengths before the black hole is formed, but this is a Lorentz-frame-dependent statement. One must be willing to give up local Lorentz invariance in the formulation of quantum gravity to argue for such coupling.

Finally, consider the transformation of the SCSET from the static frame to a frame freely falling from rest at infinity. The velocity of this frame with respect to the static frame is $v = -\sqrt{x}$. Making use of $Z_s(x)$ as defined in Eq. (3.1), the relevant physical components in the static frame can be written as

$$E = E^{\text{reg}} - F = 2(1-x)Z_s - P_r^{\text{reg}} - F, \quad P_r = P_r^{\text{reg}} - F. \tag{3.8}$$

The Lorentz transformation to the free fall frame is

$$E^{\text{ff}} = (1-x)^{-1}\left(E + xP_r + 2\sqrt{x}F\right), \quad P_r^{\text{ff}} = (1-x)^{-1}\left(xE + P_r + 2\sqrt{x}F\right),$$
$$F^{\text{ff}} = (1-x)^{-1}\left[\sqrt{x}(E + P_r) + (1+x)F\right]. \tag{3.9}$$

Then

$$E^{\text{ff}} = 2Z_s - P_r^{\text{reg}} - \frac{L_{\text{H}}}{4\pi r^2 \left(1+\sqrt{x}\right)^2}, \quad P_r^{\text{ff}} = 2xZ_s + P_r^{\text{reg}} - \frac{L_{\text{H}}}{4\pi r^2 \left(1+\sqrt{x}\right)^2},$$
$$F^{\text{ff}} = 2\sqrt{x}\, Z_s + \frac{L_{\text{H}}}{4\pi r^2 \left(1+\sqrt{x}\right)^2}. \tag{3.10}$$

The "outgoing" part of the energy flux in the free fall frame is

$$\left(F^{\text{ff}}\right)^{\text{out}} = \frac{1}{4}\left(E^{\text{ff}} + P_r^{\text{ff}} + 2F^{\text{ff}}\right) = \frac{1}{2}\left(1+\sqrt{x}\right)^2 Z_s, \tag{3.11}$$

and at the horizon is positive for spin 0, negative for spin 1. It is redshifted away to nothing in the static frame at the horizon. The "ingoing" part in the free fall frame is

$$\left(F^{\text{ff}}\right)^{\text{in}} = F^{\text{ff}} - \left(F^{\text{ff}}\right)^{\text{out}} = \frac{L_{\text{H}}}{4\pi r^2 \left(1+\sqrt{x}\right)^2} - \frac{1}{2}\left(1-\sqrt{x}\right)^2 Z_s. \tag{3.12}$$

## IV. SEMI-CLASSICAL BACKREACTION ON THE GEOMETRY

The expectation value of the renormalized SCSET can be inserted as a source in the classical Einstein equations to calculate first-order corrections to the classical spacetime geometry on which the calculation of the SCSET was based. This is not justified in all circumstances. One can imagine a "Schrodinger cat" quantum state that leads to a superposition of macroscopically different alternative geometries, rather than small fluctuations about a single classical history. However, in the evaporation of a large black hole, because a significant change in the geometry requires emission of an enormous number of Hawking quanta, and because the main effect is just a gradual decrease in the black hole's mass, semi-classical evolution does seem to make sense initially. Approaching the Page time the backreaction is substantial, and it is no longer possible to consider the backreaction as a small perturbation of a single classical history.[36] The full quantum state then consists of multiple classical histories, but new quantum fluctuations about each of them are still small, so barring new physical effects associated with the entanglement entropy saturating the Bekenstein-Hawking entropy bound, it might be argued that the semi-classical backreaction is still valid for the evolution along each of those classical histories.

To solve the Einstein equations with the SCSET source I work in advanced Eddington-Finkelstein coordinates $(v,r)$ with advanced time $v$ constant along ingoing radial null geodesics. A general spherically symmetric form of the metric in these coordinates is

$$ds^2 = -Ae^{2\psi}dv^2 + 2e^{\psi}dvdr + r^2\left(d\theta^2 + \sin^2\theta\, d\varphi^2\right), \tag{4.1}$$

following the notation of Bardeen[20]. The inverse metric has

$$g^{vv} = 0, \quad g^{vr} = e^{-\psi}, \quad g^{rr} = A \equiv 1 - \frac{2m}{r}. \tag{4.2}$$

The Einstein equations are then extremely simple, with

$$\left(\frac{\partial m}{\partial v}\right)_r = 4\pi r^2 T_v^r, \quad \left(\frac{\partial m}{\partial r}\right)_v = -4\pi r^2 T_v^v, \quad \left(\frac{\partial \psi}{\partial r}\right)_v = 4\pi r e^{\psi} T_r^v. \tag{4.3}$$

Given these components, $T_r^r \equiv T_v^v + Ae^{\psi}T_r^v$ and $T_\theta^\theta$ follows from the momentum constraint equation $T_{r;\mu}^\mu = 0$.

The $v,r$ coordinate components of the SCSET in the Schwarzschild background, where $\psi = 0$ and $A = 1 - 2M/r$, can be written in terms of the static frame components as

$$T_v^v = -E - F = -E^{\text{reg}}, \tag{4.4}$$

$$T_r^r = P_r + F = P_r^{\text{reg}}, \quad T_\theta^\theta = T_\varphi^\varphi = P_t, \tag{4.5}$$

$$T_v^r = -(1-x)F = -\frac{3}{8}P_0 x^2 k_s, \tag{4.6}$$

$$T_r^v = (1-x)^{-1}(E + P_r + 2F) = (1-x)^{-1}\left(E^{\text{reg}} + P_r^{\text{reg}}\right) = 2Z_s. \tag{4.7}$$

All are perfectly finite and smooth at $x = 1$.

It follows immediately from Eq. (4.6) and the expression for $\partial m/\partial v$ in Eq. (4.3) that $\partial m/\partial v$ is the same at all radii. With $E^{\text{reg}}$ falling off asymptotically as $(3/4)k_s P_0 x^2$, it would seem from the initial value equation giving $\partial m/\partial r$ that $m$ should diverge linearly as $r \to \infty$, but as noted earlier this is an illusion. The asymptotic contribution to $m$ is just the energy of previously emitted Hawking radiation, but this is a finite amount of energy, since the black hole was formed at a finite time in the past. For the same reason, there is a cutoff to the logarithmic divergence in the radial integral for the metric function $\psi$. The geometry stays Schwarzschild to a very good approximation in the vicinity of the black hole, with corrections of order $m_p^2/M^2$, but with a gradually decreasing gravitational mass.

There is an important missing piece of the semi-classical evolution of the black hole. We have not considered the part of the SCSET associated with quantum fluctuations of the gravitational field. The only available results are the spin 2 Hawking luminosity and the spin 2 trace anomaly. Since the spin 2 trace anomaly, with $q_2 = 212$, is more than an order of magnitude larger than the spin 1 trace anomaly, the quantum gravity contribution to the SCSET should overwhelmingly dominate that from lower spin fields, even though the spin 2 Hawking luminosity is much smaller than that of lower spins. Furthermore, general relativity is not conformally invariant, so there should be additional contributions to the trace of the SCSET not associated with the trace anomaly. It is possible that the quantum gravity SCSET is qualitatively, as well as quantitatively, different from those of ordinary quantum fields. The toy model discussed below is based on such a conjecture.

## V. IMPLICATIONS FOR THE INFORMATION PARADOX

The black hole information paradox is usually stated as a conflict between the demands of quantum theory (unitary evolution with pure states evolving into pure states, monogamy of entanglement, and locality at least in the sense of causal propagation of quantum information), and the semi-classical evaporation of black holes. Hawking quanta are entangled with "Hawking partners" inside the black hole, and as the evaporation proceeds the black hole traps increasing amounts of quantum information. Unless this quantum information can somehow escape before the black hole evaporates completely, which would seem to require acausal propagation, the result would be a mixed state.[37] A Planck scale remnant containing the enormous amount of quantum information trapped in the evaporation of a large black hole is not an attractive prospect for a number of reasons. Unruh and Wald[38] have recently reiterated the rather compelling arguments why the standard picture of black hole evaporation should lead to a mixed quantum state in the exterior of the black hole, not a pure state, and argue that this is not a violation of fundamental principles of quantum field theory, but in fact is the natural consequence of quantum field theory, even if the black hole evaporates completely without releasing its quantum information[39]. Still, there is a widely held belief, based in part on AdS-CFT, that there is a real paradox, and many more or less exotic schemes have been proposed for how the trapped information may be able to escape[40].

Black hole complementarity[22] tried to argue that all quantum information on its way into the black hole is copied onto the event horizon and then gradually leak out to infinity as subtle correlations in the apparently thermal Hawking emission, thus restoring a pure state for an external observer. The no-cloning theorem of quantum mechanics is not violated, it was claimed, because no single observer can detect both copies of the quantum information.[41] More recently Almheiri, et al (AMPS)[42] have shown that substantial entanglement of the late Hawking radiation with the early Hawking radiation, by monogamy of entanglement, means that there cannot be the entanglement of Hawking particles with Hawking "partners that makes it possible to sustain the standard semi-classical non-singular structure of the horizon as seen by a freely falling observer. This suggests that an observer freely falling across the horizon would be incinerated by a "firewall" of very high energy excitations.

The AMPS paper generated a firestorm of controversial proposals in the literature, which I will make no effort to discuss here. For a discussion of some of these see a review by Polchinski.[43] However, as long as quantum field theory and quantum gravity only allow a causal flow of quantum information on macroscopic scales I see no way that quantum information stored on the event horizon is a plausible solution to the black hole information problem. The basic issue is that the location of the event horizon for a non-quiescent black hole cannot be assumed to be always close to the apparent horizon and depends on the whole future history of the black hole. Consider a black hole formed by gravitational collapse that is quiescent for a long time. Then a null shell with 100 times the mass of the initial black hole comes in from $\Im^-$. The event horizon starts expanding in radius in anticipation of the shell's arrival, without any causal influence from the shell. How is the quantum information to know to follow the event horizon, rather than a null hypersurface closer to the apparent horizon? If the latter, it ends up well inside the event horizon and the new apparent horizon following arrival of the shell.

Furthermore, the energy conservation objections to pair creation very close to the horizon apply with even greater force to the creation of a firewall. Energetically the only way to have a firewall is to assume it is present at past null infinity, before the black hole has even formed. While there may be some quantum states with this feature, these are not physically acceptable as states in our universe, as noted by Page.[44]

Hawking, et al[45] have identified a kind of quantum mechanical "soft hair" associated with black hole event horizons, soft photons and gravitons carrying zero energy and associated with an infinite degeneracy of the vacuum. Could this "soft hair" preserve the quantum information associated with accreting matter and the generation of Hawking radiation, which would eventually leak out as subtle correlations in the Hawking radiation? Mirbabayi and Porrati[46] and Bousso and Porrati[47] argue that this soft hair is trivial and inherently incapable of carrying any quantum information, but the importance of soft hair has recently been defended by Strominger.[48] Regardless of how these issues are resolved, I do not see how soft hair can retrieve quantum information that is lost inside the horizon of a black hole. Soft hair lives at essentially infinite radius in an asymptotically flat spacetime, and has almost nothing to do with what is going on at the black hole horizon.

The Bekenstein-Hawking entropy $S_{BH} = A/(4m_p^2)$, where $A$ is the area of the horizon, $16\pi M^2$ for Schwarzschild, has an interpretation as the classical coarse-grained thermodynamic entropy of a black hole. An interpretation as the total number of quantum degrees of freedom of a black hole has been verified for certain black holes with degenerate or nearly degenerate horizons,[49] but it is not clear whether it is the microscopic entropy for more general black holes. See Wald[50] for a review. $S_{BH}$ is plausibly the maximum possible number of quantum degrees of freedom required to assemble a black hole in one operation. Quanta must have a wavelength $\lambda < \lambda_c$ and energy $\varepsilon > \varepsilon_c$, $\lambda_c \approx M$ and $\varepsilon_c \approx m_p^2/M$, in order to be captured by the black hole. The maximum number forming the black hole is then roughly $M/\varepsilon_c \approx M^2/m_p^2 \approx S_{BH}$. However, suppose the black hole cycles through stages of accretion and of Hawking evaporation. The number of quantum degrees of freedom stored in the black hole would seem to grow indefinitely, unless quantum information can be removed (acausally) from the black interior.

I would argue a black hole is in general not a conventional quantum system with a fixed number of degrees of freedom proportional to its surface area, and a young black hole formed by stellar collapse is really just a rather empty region of spacetime. A physically more appropriate microscopic measure of its entropy is the *entanglement* (von Neumann) entropy $S_{vN}$ of the black hole as a subsystem of the fields on a Cauchy hypersurface, renormalized so as *not* to include the short-range correlations of the vacuum across the horizon that are present across any sharp boundary. This is the total number of degrees of freedom in the interior entangled with the exterior of the black hole, arising from the initial formation of the black hole, any subsequent accretion, and the entanglement generated by the emission of Hawking radiation. For a young black hole $S_{vN}$ is typically tiny compared with $S_{BH}$, since a star whose collapse forms the black hole is made up of quanta with energies $\varepsilon \gg \varepsilon_c$. Emission of Hawking radiation causes $S_{vN}$ to increase, and in the absence of other influences it equals $S_{BH}$ at the Page time[51], when the black hole has lost about 1/2 of its original mass.

It is at the Page time that one is really forced to deal with the black hole information problem. There is nothing in the usual semi-classical theory of black hole evaporation that would explain why Hawking evaporation should stop at the Page time. If Hawking radiation continues to be emitted, and there is no way of retrieving the quantum information from deep inside the black hole, the ratio $S_{vN}/S_{BH}$ would continue to increase and eventually become much larger than one.

My approach is inspired by the Bousso covariant entropy bound conjecture.[23] It states that the entropy flux crossing a light sheet generated by non-diverging null geodesics orthogonal to a spacelike two-surface cannot exceed 1/4 the area of the two-surface. The light sheet extends only as far as there are no caustics. The conjecture was proven by Flanagan, et al[52] in the context of classical physics, where the entropy flux can be expressed in terms of an energy-momentum tensor satisfying the null energy condition (NEC), $k^\alpha k^\beta T_{\alpha\beta} \geq 0$ for any null vector $k^\alpha$. In the context of quantum fields in curved spacetimes the von Neumann entropy flux cannot be expressed in terms of the energy-momentum tensor and the NEC may be violated locally.

A quantum version of the Bousso bound in which the entanglement entropy across the two-surface is added to the area was suggested by Strominger and Thompson.[53] This approach has been refined by Bousso, et al,[21] who define a generalized entropy $S_{gen}(\sigma)$ for a compact 2-surface $\sigma$ with area $A(\sigma)$ dividing a Cauchy hypersurface into two regions. With $S_{out}(\sigma)$ the von Neumann entropy of the "external" non-compact region,

$$S_{\text{gen}}(\sigma) = S_{\text{out}}(\sigma) + \frac{A(\sigma)}{4m_{\text{p}}^2}. \tag{5.1}$$

This generalized entropy is used to formulate a "quantum focusing conjecture" (QFC). Deforming $\sigma$ along null geodesics orthogonal to $\sigma$ gives compact 2-surfaces $\sigma'$. For $\sigma'$ that are uniform affine distance $\lambda$ from $\sigma$, the QFC implies that $d^2 S_{\text{gen}}(\sigma')/d\lambda^2 \leq 0$, so if $dS_{\text{gen}}/d\lambda \leq 0$ initially, it remains non-positive until the null hypersurface (light sheet) hits a singularity or has a caustic. The QFC implies the "quantum Bousso bound"

$$S(\sigma') - S(\sigma) \leq \frac{[A(\sigma) - A(\sigma')]}{4m_{\text{p}}^2} \tag{5.2}$$

and implies a "quantum singularity theorem" analogous to the Penrose singularity theorem in classical theory. If the QFC is true, it would seem to preclude any release of quantum information from the interior of a black hole. I will comment on evidence against the QFC as a general principle in VII.

What I speculate might be a property of quantum gravity (but not generic quantum fields) is a different entropy bound, that with $S(\sigma)$ and $A(\sigma)$ as defined above

$$S(\sigma) - S(\sigma') \leq \frac{[A(\sigma) - A(\sigma')]}{4m_{\text{p}}^2}, \tag{5.3}$$

i.e., that the entanglement entropy on a non-expanding light sheet from $\sigma$ to $\sigma'$ is bounded by the area difference. Support for this "quantum gravity conjecture" (QGC) is provided by AdS-CFT, where Ryu and Takayanagi[54] have shown that entanglement in the CFT between disjoint parts of the boundary of AdS has as a bulk dual an Einstein-Rosen bridge, with the entanglement entropy equal to 1/4 of its minimal area in Planck units. An explicit example was worked out by Jensen, et al.[55] The QGC is consistent with the ER=EPR conjecture of Maldacena and Susskind.[56] If microscopic Einstein-Rosen bridges connect entangled qbits, it is reasonable that there is a large backreaction modifying the macroscopic geometry when these approach a density of one per Planck area across a two-surface. Such a backreaction could conceivably enforce the QGC.

Consider the implications of the QGC for the evolution of the interior of a spherically symmetric black hole, assuming that quantum backreaction prevents formation of the classical $r = 0$ singularity and that a macroscopic spacetime geometry provides an arena for the quantum effects. The absence of a singularity means there must be an inner apparent horizon, so $r = 0$ can be a regular point on spacelike slices. Any quantum information entering the black hole will end up very close to the inner apparent horizon on a dynamical advanced time scale of a few times $M$. Apply the QGC to the light sheet generated by *ingoing* radial geodesics from a spherical 2-surface just outside the inner apparent horizon, at which the entanglement entropy $S$ is close to the entanglement entropy of the outer apparent horizon at at the same advanced time. While the radius of the inner apparent horizon $r_-$ may be very small compared with $M$ initially, as the entanglement entropy $S_{vN}$ of the black hole grows over time $r_-$ must increase and approach the radius of the outer apparent horizon by the Page time. As long as the geometry is close to Schwarzschild in the vicinity of the outer apparent horizon it seems reasonable that the standard semi-classical estimates for the Hawking luminosity and the increase in $S_{vN}$ should be applicable, but absent a full theory of quantum gravity the ultimate fate of the black hole is wide open to speculation.

A toy model for the evolution of the metric in this scenario, and the properties of the energy-momentum tensor related to the metric through the classical Einstein equations, is discussed in Part VI.

## VI. A TOY MODEL FOR BLACK HOLE EVAPORATION

I assume the general relativistic structure of spacetime survives in the interior of a large black hole. Quantum fluctuations generate an effective stress-energy tensor that is related to the metric of an approximate "quasi-classical" background spacetime by the Einstein equations. The only reason this has a chance of making sense when the quantum backreaction is large is the enormous entropy, so that any one mode of the quantum fields is a tiny perturbation on the bulk geometry. The toy model is meant to describe a large spherically symmetric black hole long after formation, but well before the Page time, with evolution driven by Hawking radiation. The ansatz for the metric and its evolution is motivated by the QGC of Part V, but is quite ad hoc in its details.

The two metric functions $m(v,r)$ and $\psi(v,r)$ of a general spherically symmetric metric in advanced Eddington-Finkelstein coordinates, Eq. (4.1), are assumed to have the form

$$m = \frac{Mr^3}{r^3 + 2Ma^2}, \quad e^\psi = \frac{r^6 + (ca)^6}{r^6 + (2Ma^2)^2}. \tag{6.1}$$

In the interior of the black hole $M$ and $a$ are functions of the advanced time $v$, chosen to accommodate a radial inflow of negative energy and entanglement associated with emission of Hawking radiation and maintain the QGC entropy bound of Eq. (5.3). As long as $2M/a \gg 1$ the geometry is close to Schwarzschild for $r^3 \gg 2Ma^2$ and approaches deSitter as $r \to 0$. There is an outer apparent horizon at $r_+ \cong 2M$ and an inner apparent horizon at $r_- \cong a$. The apparent horizons become degenerate and disappear if and when $2M/a \leq 3\sqrt{3}/2$. Nonsingular black hole models proposed by Hayward[57] and Frolov[58] have the same form for $m$, but with $a$ a constant the order of the Planck length, and $e^\psi = 1$. A similar form for $m$ was suggested by Bonanno and Reuter,[59] based on a renormalization group approach to quantum gravity.

A previous version of my toy model[26] also had $e^\psi = 1$. Then the surface gravity of the inner apparent horizon $\kappa_i \cong -1/a$ is much greater in magnitude than the surface gravity of the outer horizon, $\kappa_o \cong 1/(4M)$. The negative sign of $\kappa_i$ means there is an exponentially growing *blue* shift associated with the inner apparent horizon, instead of the red shift associated with the outer apparent horizon, a kind of instability it seems desirable to minimize. With the $e^\psi$ of Eq, (6.1), $\kappa_i \cong -e^\psi / a \cong -c^6 a / (2M)^2$. A somewhat similar "redshift function" was introduced recently by Frolov and Zel'nikov[60].

All disturbances propagating on null geodesics inside the outer apparent horizon end up converging on the inner apparent horizon within a few dynamical times, so almost all of the quantum information acquired by the black hole should end up there. The entanglement entropy $S(\sigma)$ of a sphere a bit outside the inner apparent horizon should be close to the entanglement entropy $S_{vN}$ of the black hole at a given advanced time. A light sheet generated by ingoing radial null geodesics from this sphere and ending at a sphere $\sigma'$ well inside the inner apparent horizon has $S(\sigma') \ll S(\sigma)$ and $A(\sigma') \ll A(\sigma)$. As the black hole evaporates, the implication of the QGC bound in the context of the toy model is that $a(v)$ must increase to keep $S(\sigma) \leq \pi a^2 / m_P^2$.

Page[51] has estimated that the rate of increase of the von Neumann entropy associated with the standard semi-classical Hawking luminosity of photons and gravitons is

$$\frac{dS_{vN}}{dv} \approx \frac{1}{715M}, \tag{6.2}$$

from which, assuming the QGC bound is saturated,

$$\frac{da^2}{dv} \approx \frac{1}{2250} \frac{m_P^2}{M} \equiv 2\beta\pi(2M)^3 \sigma T_H^4, \tag{6.3}$$

where $\beta \approx 21.5$. The Hawking luminosity gives directly

$$\frac{dM}{dv} \approx -L_H = -\alpha\pi(2M)^2 \sigma T_H^4, \tag{6.4}$$

$\alpha = \sum_s k_s$.

The stress-energy tensor (SET) components implied by the metric of Eq. (6.1) are found from Eqs. (4.3) and the momentum conservation equation, with the result, if $M$ and $a^2$ depend only on the advanced time $v$ (valid only in the black hole interior),

$$T_v^v = -\frac{3}{8\pi} \frac{a^2 (2M)^2}{(r^3 + 2Ma^2)^2}, \quad e^\psi T_r^v = \frac{3r^4}{2\pi} \frac{(2Ma^2)^2 - (ca)^6}{[r^6 + (ca)^6][r^6 + (2Ma^2)^2]}, \tag{6.5}$$

$$T_v^r = -\frac{1}{4} r(2M)^2 \sigma T_H^4 \frac{r^3 \alpha + (2M)^3 \beta}{(r^3 + 2Ma^2)^2}, \tag{6.6}$$

and

$$2T_\theta^\theta = e^{-\psi}\left(re^\psi T_r^v\right)_{,v} + \frac{1}{r}\left(r^2 T_v^v\right)_{,r} + \frac{1}{r}\left(r^2 A e^\psi T_r^v\right)_{,r} + \left(\frac{1}{2}A_{,r} + A\psi_{,r}\right) re^\psi T_r^v. \tag{6.7}$$

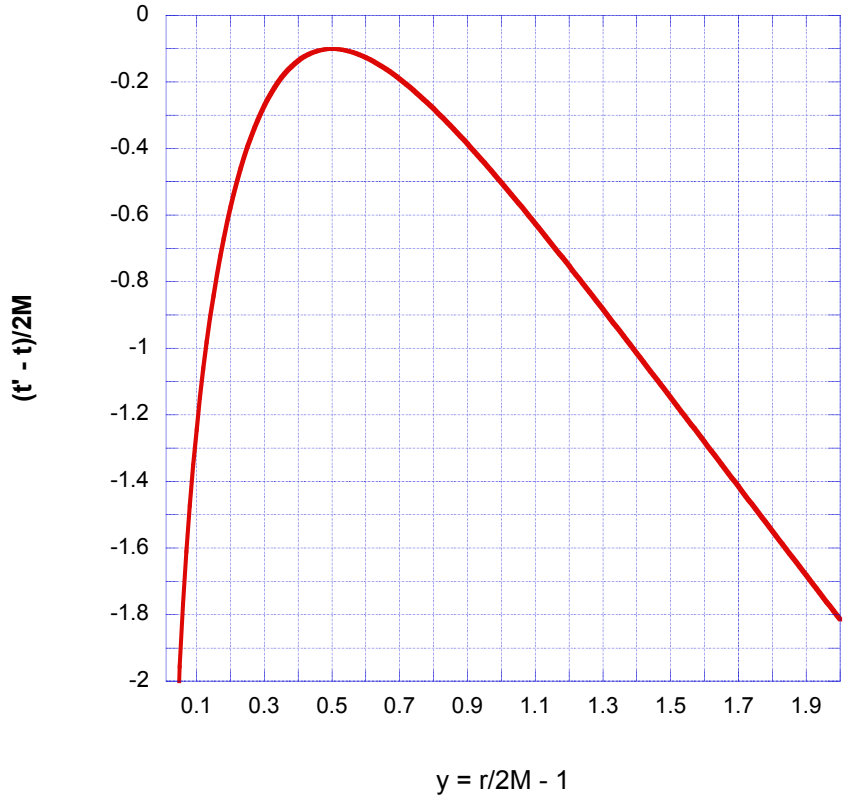

Figure 9. The difference of $t'$ from the Schwarzschild time coordinate $t$ between $r = 2M$ and $r = 4M$, ignoring the time dependence of $M$, for $t'$ as defined by Eq. (6.8).

In order to extend the model to the black hole exterior, $M$ and $a$ should be considered functions of a time coordinate $t'$ that smoothly transitions over a range of $r$ from the advanced time $v$ in the interior to a retarded time well outside the black hole. The idea is that outside the black hole the constant-$t'$ hypersurface should track the energy and quantum information carried away from the black hole by the outgoing Hawking radiation. With $y \equiv (r/2M) - 1$ and for $a \ll M$, a suitable $t'(v,r)$ for $y > 0$ is

$$t' = v - 4M(t')\left[y + \frac{1}{2}\ln(1+4y^2) - \frac{1}{2}\tan^{-1}(2y)\right]. \tag{6.8}$$

In practice, since the evaporation time scale is very large compared with $M$ and $T_v^r$ is tiny compared with the other components of the SET for $a \gg m_p$, $M$ in Eq. (6.8) can be considered a constant for $y = O(1)$. The time $t'$ is close to the Schwarzschild time $t = v - r - 2M\ln(r/2M - 1)$ around $r = 3M$, as can be seen from Fig. 9.

The components of the model SET in the core of the black hole, except for $T_v^r$, which is negligible compared to the others as long as $m_p^2/(2Ma) \ll 1$, are plotted in Fig. 10 for $2M/a = 1000$ and $c = 1.2$. The geometry approaches Schwarzschild, and the SET starts falling off rapidly once $r/a > (2M/a)^{1/3} = 10$. The power law falloff does mean that, while the SET is small at the outer apparent horizon, there is quantum hair in the exterior of the black hole that, for $a \gg m_p$, far exceeds the semi-classical result for spin 0 and spin 1 fields, and in principle could be used by an external observer to measure the entanglement entropy of the black hole. While this may seem rather strange in the context of conventional quantum mechanics, it does not obviously violate any fundamental principles.

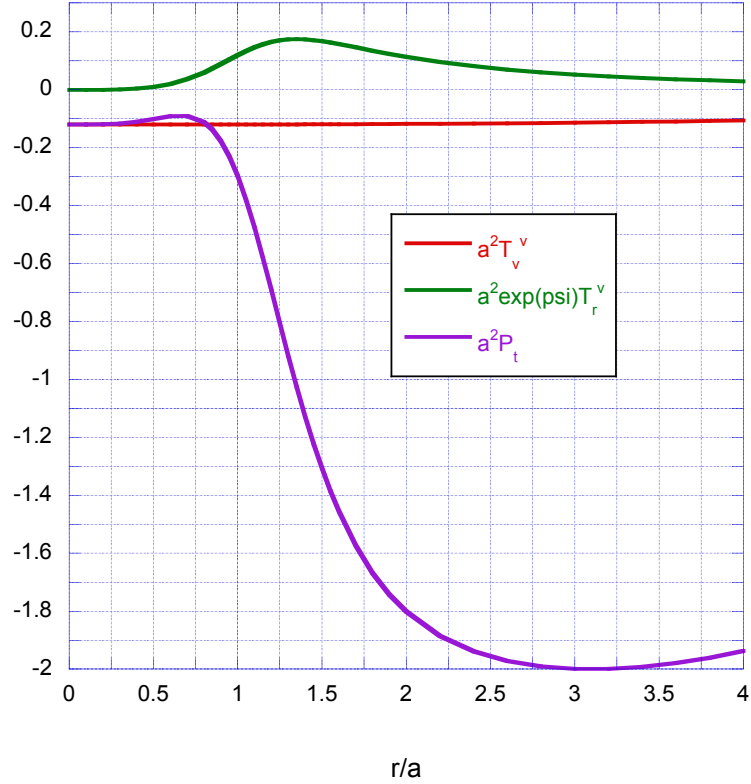

Figure 10. The dominant components of the toy model SET are plotted for $2M/a = 1000$ and $c = 1.2$ over a range of radii in the core of the black hole.

Consider null geodesics in the toy model geometry. The angular momentum $L = k_\varphi$ is a constant of the motion, and $k_v$ is approximately constant on dynamical time scales, though not on evaporation time scales. The null condition is

$$2e^{-\psi} k_r k_v + A k_r^2 + \frac{L^2}{r^2} = 0, \qquad (6.9)$$

from which

$$k^r = \pm\sqrt{e^{-2\psi} k_v^2 - L^2/r^2}. \qquad (6.10)$$

The condition that the geodesic be future-directed is

$$0 \leq e^{\psi} k^v = k_r = \frac{k^r - e^{-\psi} k_v}{A}. \qquad (6.11)$$

Where $A > 0$ outside the outer apparent horizon and inside the inner apparent horizon, this requires $k_v < 0$. Where $A < 0$, between the two black hole apparent horizons, only the negative sign in Eq. (6.10), but both signs of $k_v$, are allowed. Ingoing geodesics with $k_v < 0$ can easily cross both the outer and inner apparent horizons, and for these it is convenient to write

$$k_r = \frac{L^2}{r^2} \frac{1}{\left(-e^{-\psi} k_v + \sqrt{e^{-2\psi} k_v^2 - AL^2/r^2}\right)}. \tag{6.12}$$

Positive $k_v$ means (neglecting the time variation of the metric) a particle on the trajectory has negative Killing energy relative to infinity, while future-directed means positive energy relative to a local time-like observer. This is the usual concept of a Hawking "partner" in the local pair creation model of Hawking radiation that I have argued is inconsistent with the SCSET and local conservation of energy. For a metric independent of $v$, such a trajectory approaches but cannot cross the inner apparent horizon. The geodesic equation, allowing for time variation, gives

$$\frac{dk_v}{d\lambda} = \frac{m_{,v}}{r} k_r^2 + \psi_{,v} e^{-\psi} k_v k_r, \tag{6.13}$$

or

$$\frac{d}{d\lambda}\left(e^{-\psi} k_v\right) = e^{-\psi} \frac{m_{,v}}{r} k_r^2 - e^{-\psi} k_v k^r \psi_{,r}. \tag{6.14}$$

An "outward" radial null geodesic, with $L = 0$, and originating between the apparent horizons with $k_v > 0$, has $k^r = -e^{-\psi} k_v$ and $k_r = -2e^{-\psi} k_v / A$. As it approaches the inner apparent horizon the first term in Eq. (6.14) dominates, and since $m_{,v} < 0$, $e^{-\psi} k_v$ and $k^r$ are both forced to zero as $A \to 0$. The trajectory crosses the inner apparent horizon and $e^{-\psi} k_v$ and $dr/d\lambda$ simultaneously change sign, while $k_r$ stays positive and finite. If $L \neq 0$, $k_v$ can become negative while $A < 0$, and then $dr/d\lambda$ stays negative across the inner apparent horizon, but will change sign not far inside. While the outer apparent horizon of an evaporating black hole is a timelike hypersurface, the inner apparent horizon is a spacelike hypersurface as long as its radius keeps increasing. An ingoing null geodesic inside the inner apparent horizon becomes outgoing at a turning point (or passes through the origin if $L = 0$) and approaches, but cannot cross, the inner apparent horizon from the inside.

This behavior of geodesics in the toy model suggests the presence of a large outward flow of energy just inside the inner apparent horizon, corresponding to a large value of the $T_r^v$ component of the SET, beyond that implied by the toy model metric, in a sense a kind of firewall. However, the actual physics of the evolution inside the inner apparent horizon is presumably quite complicated. The toy model does not take into account what happens to the collapsing star that formed the black hole, and whose remains must still be present inside the inner apparent horizon and presumably will interact strongly with the degrees of freedom entangled with the Hawking radiation. The toy model should only be considered a demonstration of the *possibility* of a non-singular evolution of the interior of the black hole, and should not be taken too seriously in its details. In particular, an increasing flow of energy along the inner apparent horizon will increase $(\partial \psi / \partial r)_v$ and make $e^\psi$ even smaller in the core of the black hole, further suppressing the magnitude of $\kappa_i$ and the growth rate of the blue shift along the inner apparent horizon. The blue shift is related to the "mass inflation" instability of Poisson and Israel,[61] though by construction in the toy model the mass function $m \cong a/2$ at the inner apparent horizon only increases slowly with advanced time.

The energy density measured in the local frame of an observer freely falling from outside the black hole as she crosses either apparent horizon can be expressed simply in terms of the coordinate components as given in Eqs. (6.5) and (6.6). Neglecting the slow time variation of the metric, the 4-velocity $u^\alpha = dx^\alpha / d\tau$ of the observer at a point where $A = 0$ is

$$u_v = -\varepsilon, \ u_r = \frac{1}{2}e^\psi \varepsilon, \ u^v = \frac{1}{2}\varepsilon, \ u^r = -e^{-\psi}\varepsilon, \tag{6.15}$$

with $\varepsilon > 0$ and $\varepsilon = 1$ for free fall from infinity. Then the energy density is

$$T^{(0)(0)} = u^\alpha T_\alpha^\beta u_\beta = \varepsilon^2 \left( -T_v^v + \frac{1}{4}e^\psi T_v^r + e^{-\psi}T_r^v \right), \tag{6.16}$$

which is positive for the toy model at both apparent horizons.

A classical SET is normally assumed to obey the local null energy condition $T_{\alpha\beta}k^\alpha k^\beta \geq 0$ (NEC) for any null geodesic tangent vector $k^\alpha = dx^\alpha / d\lambda$, but this is violated both by the SCSET and the toy model SET. Using Eqs. (6.10) and (6.11), the null energy can be written

$$T_{\alpha\beta}k^\alpha k^\beta = e^\psi T_r^v \left(k^r\right)^2 + e^{-\psi}T_v^r k_r^2 + \left(-T_v^v + T_\varphi^\varphi\right)\frac{L^2}{r^2}. \tag{6.17}$$

The sign of the SCSET $T_r^v$ is the sign of $Z_s$ (see Eq. (4.7)), positive for spin 0 and negative for spin 1 at the horizon, while in the toy model Eq. (6.5) shows that $T_r^v > 0$ everywhere as long as $c < (2M/a)^{1/3}$. The second term is negative for both the SCSET and the toy model, but in the toy model is relatively small except where $e^{-\psi} \gg 1$ in the inner core. In the third term, $-T_v^v > 0$ everywhere in the toy model, but over much of the range $a < r < (2Ma^2)^{1/3}$ the tangential stress is negative and $-T_v^v + T_\varphi^\varphi < 0$. At larger radii the tangential stress becomes positive, but starts decreasing rapidly.

One suggestion for a replacement for the NEC in the context of quantum field theory is the *averaged* null energy condition (ANEC)

$$\int d\lambda T_{\alpha\beta}k^\alpha k^\beta \geq 0, \tag{6.18}$$

in which the integral extends over every complete null geodesic. This less restrictive energy condition has been proven under fairly general assumptions in flat, but not curved, spacetimes (see Faulkner, et al[62]). For the SCSETs of Part II, it is easy to check the ANEC on the (unstable) circular null geodesic at $r = 3M$ with the results for the Unruh state of conformally coupled quantum fields that it is satisfied by spin 0, but violated by spin 1. Levi and Ori[33] showed that the ANEC is violated on the circular null geodesic for a spin 0 minimally coupled quantum field. The toy model probably violates the ANEC, but no definite conclusion is possible because the scope of the toy model does not extend to the evolution of the entire spacetime. Flanagan[63] has established a quantum null energy condition in 2D curved spacetimes, but his proofs cannot be extended to 4D.

Is the QFC of Bousso, et al.[23] mentioned in Part V a viable alternative? The problem is that it is inconsistent with the semi-classical analysis of black hole evaporation in a Schwarzschild background as described in Part IV. Consider a light sheet constructed from "outgoing" null geodesics from a sphere $\sigma$ at radius $r_0$ just slightly inside the *apparent* horizon at $r = 2M$ in the Schwarzschild geometry. As long as the light sheet is very close to the apparent horizon, the entanglement entropy across the light sheet should increase as given by Eq. (6.2). The radius of the light sheet as a function of advanced time is approximately

$$r \cong r_0 - \varepsilon v - 4\pi M T_v^r v^2 \cong r_0 - \varepsilon v + L_H \frac{v^2}{4M}, \tag{6.19}$$

with $\varepsilon \cong (2M - r_0)/4M \ll 1$. For a large black hole emitting only photons and gravitons, then, Eq. (1.2) gives

$$\frac{dS_{\text{gen}}}{dv} \cong \frac{4\pi M}{m_p^2} \left[ \frac{m_p^2}{715\pi (2M)^2} - \varepsilon + \frac{7.23 m_p^2}{61440\pi (2M)^2} \frac{v}{2M} \right]. \tag{6.20}$$

For a wide range of values of $\varepsilon$, $(m_p/M)^2 \ll \varepsilon \ll m_p/M$, the approximations behind Eq. (6.19) and using the SCSET as an average over the vacuum fluctuations giving rise to Hawking radiation are both valid during the evolution from an initially negative to a positive $dS_{\text{gen}}/dv$ at $v/2M \gg 1$. Eq. (6.20) violates the QFC, which implies that there is no basis for the quantum singularity theorem of Bousso, et al[23] and for assuming the quantum Bousso entropy bound of Eq. (5.2) holds indefinitely into the future.

The black hole evolution scenario of the toy model cannot be extended all the way to the Page time, when $a/M$ becomes of order one and there are substantial deviations from Schwarzschild in the vicinity of the outer apparent horizon. Approaching the Page time, there is no justification for the evolution of $M$ and $a^2$ given by Eqs. (6.3) and (6.4) and the form of $T_v^r$ given in Eq. (6.6). The inner horizon approaches the outer horizon, but there is no guidance about the further evolution. If the horizon becomes degenerate, the surface gravity goes to zero, and naively one might expect the Hawking luminosity to go to zero. Conceivably, one could be left with a stable massive remnant black hole with a degenerate horizon that is no longer losing energy and whose von Neumann entropy equals its Bekenstein-Hawking entropy. The quantum information in the black hole would remain trapped indefinitely for external observers. This outcome is not completely satisfactory because it suggests the presence of a Cauchy horizon, implying a breakdown of predictability. Geodesics inside the degenerate horizon reach infinite advanced time at a finite affine parameter.

A second possibility is that the horizons become degenerate, but quantum fluctuations in the geometry allow the quantum information and energy propagating just inside the degenerate horizon to gradually leak out, with the black hole slowly shrinking down to the Planck scale and then disappearing completely without any loss of quantum information.

A third possibility is that the inner and outer apparent horizons merge and disappear around the Page time. The absence of trapped surfaces would allow all of the trapped energy and quantum information to escape suddenly in a burst of radiation.

The final resolution of these issues awaits a full theory of quantum gravity.

## VII. SUMMARY

The semi-classical stress-energy tensors for conformally coupled spin 0 and spin 1 fields in the exterior spacetime of a large Schwarzschild black hole were calculated in the 1980's and 1990's, but the results were not always presented in a way that facilitated their physical interpretation. I have tried to remedy that in this paper. I believe these calculations need to be taken very seriously in addressing the black hole information problem, since for a large astrophysical black hole the semi-classical approximation is the first order of an expansion in powers of an incredibly small expansion parameter. There is no local significance to the event horizon, and to challenge local quantum field theory in the vicinity of a large black hole would seem to also challenge the validity of local quantum field theory in Minkowski spacetime as an approximation in gravitational fields of comparable strength in laboratories on the Earth, where it has been tested to exquisite precision. I have shown that the semi-classical theory gives no indication that Hawking radiation is anything but a low energy phenomenon, associated with tidal disruption of vacuum fluctuations in the general vicinity of the black hole, and should not be considered the result of pair creation within some small Planck scale neighborhood of the horizon.

These conclusions are based on polynomial fits to the numerical data, where I improve on a previous fit by Visser for the spin 0 case. Some significant ambiguities remain for the spin 1 case, where the precise results of the earlier calculations are no longer available. A current project initiated by Levi and Ori[32] to revisit and extend the numerical results has the potential of clearing up these ambiguities as well as giving results for more rapidly evolving and only axisymmetric black holes.

My take on the black hole information problem is not new, but it has become unfashionable in certain quarters over the last 20 years or so. I see no way to prevent much and probably the great bulk of the quantum information associated with Hawking "partners" from ending up deep inside the black hole. As long as the black hole geometry is close to Schwarzschild, it cannot be retrieved without drastically acausal propagation, which would seem to be a much more serious violation of conventional quantum mechanics and quantum field theory than the failure to retrieve it. Furthermore, a Schwarzschild black hole event horizon is locally indistinguishable from a Rindler horizon in Minkowski spacetime. Quantum information can cross a Rindler horizon, without leaving behind any significant trace to be retrieved by a uniformly accelerating observer remaining outside the horizon.

Even in light of AdS-CFT correspondence, which implies the bulk quantum fields must be unitary, I see no reason why it is necessary for all the quantum information to escape to the boundary. The quantum fields in the exterior of the black hole are a subsystem, which is not expected to be in a pure state even if the total system is in a pure state. Papadodimas and Raju[64] have suggested how the boundary CFT may be able to track the evolution of the black hole interior. This requires no acausal communication, since the quantum state evolves deterministically in both the bulk and the boundary.

Still, there are rather good reasons why one would not like to let a black hole evaporate down to the Planck scale without releasing its trapped quantum information. If the Bekenstein-Hawking entropy is to have any microscopic meaning, it should be at least an upper limit to the entanglement entropy between the interior and the exterior of the black hole. In the absence of acausal propagation of quantum information, this is only possible if something happens to prevent continued evaporation past the Page time, when the entanglement entropy between the Hawking partners in the interior and the Hawking radiation in the exterior equals the Bekenstein-Hawking entropy. The large macroscopic entanglement across the horizon is what distinguishes a black hole horizon from a null hypersurface in Minkowski spacetime. As a speculation on a possible scenario, I presented a toy model in Part VI for how quantum backreaction might modify the Schwarzschild geometry, possibly enough to either to eliminate a horizon completely or at least turn off the Hawking radiation by the Page time. Such a large backreaction is far beyond what is expected from the usual SCSET, and would have to be a novel property of quantum gravity. Any observational test of these ideas depends on astronomers being lucky enough to detect Hawking radiation from a small primordial black hole that is just now reaching its Page time. Does the Hawking radiation end with a bang as the Hawking temperature keeps increasing, with a whimper as the Hawking temperature goes to zero, or with a superbang if the black hole suddenly dissolves?

ACKNOWLEDGEMENTS
I thank Andreas Karch and Ivan Muzinich for a number of helpful conversations, and Don Page for encouragement and help with the early literature. Also, I am very grateful to the Perimeter Institute for hosting a number of very stimulating visits that exposed me to a wide range of views on this and related topics, and during which some of the research for and writing of this paper was done.

REFERENCES

[1] S. W. Hawking, Nature **248,** 30 (1974); Commun. Math. Phys. **43,** 199 (1975).
[2] D. N. Page, Phys. Rev. D **13**, 198 (1976).
[3] T. Elster, Phys. Lett. **94A**, 205 (1983).
[4] B. E. Taylor, C. M. Chambers and W. A. Hiscock, Phys. Rev. D **58**, 044012 (1998) [arXiv:gr-qc/9801044].
[5] S. M. Christensen, Phys. Rev. D **14**, 2490 (1976).
[6] S. M. Christensen and S. A. Fulling, Phys. Rev. D **15**, 2088 (1977).
[7] J. B. Hartle and S. W. Hawking, Phys. Rev. D, **13**, 2188 (1976).
[8] W. G. Unruh, Phys. Rev. D **14**, 870 (1976).
[9] K. W. Howard and P. Candelas, Phys. Rev. Lett. **53**, 403 (1984).
[10] P. R. Anderson, W. A. Hiscock and D. A. Samuel, Phys. Rev. Lett. **70**, 1739 (1993).
[11] T. Elster, Class. Quantum Grav. **1**, 43 (1984).
[12] B. P. Jensen and A. C. Ottewill, Phys. Rev. D **39**, 1130 (1989).
[13] B. P. Jensen, J. G. McLaughlin and A. C. Ottewill, Phys. Rev. D, **43**, 4142 (1991).
[14] M. Visser, Phys. Rev. D **56**, 936 (1997) [arXiv:gr-qc/9703001].
[15] D. N. Page, Phys. Rev. D **25**, 1499 (1982).
[16] M. R. Brown and A. C. Ottewill, Phys. Rev. D **31**, 2514 (1985).
[17] M. R. Brown, A. C. Ottewill and D. N. Page, Phys. Rev. D **33**, 2840 (1986).
[18] J. Matyjasek, Class. Quantum Grav. **14**, L15 (1997).
[19] J. Matyjasek, Phys.Rev. D **55**, 809 (1997).
[20] J. M. Bardeen, Phys. Rev. Lett. **46**, 382 (1981).
[21] R. Bousso, Z. Fisher, S. Leichenauer and A. C. Wall, Phys. Rev. D **93**, 064044 (2016) [arXiv:1506.02669].
[22] L. Susskind, L. Thoracius and J. Uglum, Phys. Rev. D **48**, 3743 (1993) [arXiv:hep-th/9306069].
[23] R. Bousso, JHEP **07**, 004 (1999) [arXiv:hep-th/9905177]; Rev. Mod. Phys. **74**, 825 (2002) [arXiv:hep-th/0203101].
[24] J. M. Bardeen, [arXiv:1406.4098].
[25] M. J. Duff, Nucl. Phys. B**125**, 334 (1977).
[26] P. Candelas, Phys. Rev. D **21**, 2185 (1980).
[27] R. Simkins, Ph. D. thesis, Pennsylvania State Univ. (1986), unpublished.
[28] B. E. Taylor, C. M. Chambers and W. A. Hiscock, Phys. Rev. D **58**, 044012 (1998); [arXiv:gr-qc/9801044].
[29] V. P. Frolov and A. I. Zel'nikov, Phys. Rev. D **35**, 3031 (1985).
[30] A. Levi and A. Ori, Phys. Rev. Lett. **117**, 231101 (2016) [arXiv:1608.03806].
[31] W. G. Unruh, Phys. Rev. D **15**, 365 (1977).
[32] S. A. Fulling, Phys. Rev. D **15**, 2411 (1977).
[33] M. K. Parikh and F. Wilczek, Phys. Rev. Lett. **85**, 5042 (2000) [arXiv:hep-th/9907001].
[34] M. Visser, Int. J. Mod. Phys. D **12**, 649 (2003) [arXiv:0106111].
[35] D. Baker, D. Kodwani, U.-L. Pen and I-S. Yang, [arXiv:1701.0481].
[36] J. Hartle and T. Hertog, Phys. Rev. D **92**, 063509 (2015) [arXiv:1502.06770].
[37] S. W. Hawking, Physical Review D **14**, 2460 (1976).

[38] W. G. Unruh and R. M. Wald, Reports on Progress in Physics, to be published [arXiv:1703.02140].

[39] T. Maudlin, [arXiv:1705.03541].

[40] e.g., S. B. Giddings and Y. Shi, Phys. Rev. D **87**, 064031 (2013) [arXiv:1205.4732].

[41] L. Susskind and L. Thoracius, Phys. Rev. D **49**, 966 (1994) [arXiv:hep-th/9308100].

[42] A. Almheiri, D. Marolf, J. Polchinski and J. Sully, JHEP, 1302, 062 (2013) [arXiv:1207.3123]; A. Almheiri, D. Marolf, J. Polchinski, D. Stanford and J. Sully [arXiv:1304.6483].

[43] J. Polchinski, notes from lectures at the 2015 Jerusalem Winter School and the 2015 TASI [arXiv:1609.04036].

[44] D. N. Page, JCAP **1406**, 051 (2014) [arXiv:1306.0562].

[45] S. W. Hawking, M. J. Perry and A. Strominger, Phys. Rev. Lett. **116**, 231301 (2016) [arXiv:1601.00921].

[46] M. Mirbabayi and M. Porrati, Phys. Rev. Lett. **117**, 211301 (2016) [arXiv:1607.03120].

[47] R. Bousso and M. Porrati, [arXiv:1706.00436].

[48] A. Strominger, [arXiv1706.07143].

[49] e.g., A. Strominger and C. Vafa, Phys. Lett. B **379**, 99 (1996) [arXiv:hep-th/9601029]; A. Strominger, JHEP **9802**:009 (1998) [arXiv:hep-th/9712251].

[50] R. M. Wald, Living Rev. Rel. **4**, 6 (2001) [arXiv:gr-qc/9912119].

[51] D. N. Page, JCAP **1309**, 028 (2013) [arXiv:1301.4995].

[52] E. E. Flanagan, D. Marolf and R. M. Wald, Phys. Rev. D **62**, 084035 (2000) [arXiv:hep-th/9908070].

[53] A. Strominger and D. Thompson, Phys. Rev. D **70**, 044007 (2004) [arXiv:hep-th/0303067].

[54] S. Ryu and T. Takayanagi, Phys. Rev. Lett. **96**, 181602 (2006) [arXiv:hep-th/0603001]; JHEP **0608**, 045 (2006) [arXiv:hep-th/0605073].

[55] K. Jensen and A. Karch, Phys. Rev. Lett. **111**, 211602 (2013) [arXiv:1307.1132]; K. Jensen, A. Karch and B. Robinson, Phys. Rev. D **90**, 064019 (2014) [arXiv:1405.2065].

[56] J. Maldacena and L. Susskind, Fortsch. Phys. **61**, 781 (2013) [arXiv:1306.6483].

[57] S. A. Hayward, Phys. Rev. Lett. **96**, 031103 (2006) [arXiv:gr-qc/0506126].

[58] V. P. Frolov, JHEP **1405**, 049 (2014) [arXiv:1402.5446].

[59] A. Bonanno and M. Reuter, Phys. Rev. D **73**, 083005 (2006) [arXiv:hep-th/0602159].

[60] V.P. Frolov and A. Zel'nikov, Phys. Rev. D **95**, 044042 (2017) [arXiv:1612.05319]; see also [arXiv:1704.03043].

[61] E. Poisson and W. Israel, Phys. Rev. D **41**, 1796 (1990).

[62] T. Faulkner, R. G. Leigh, O. Parrikar and H. Wang, JHEP **1609**, 038 (2016) [arXiv:1605.08072].

[63] E. E. Flanagan, Phys. Rev. D **66**, 104007 (2002) [arXiv:gr-qc/0208066].

[64] K. Papadodimas and S. Raju, Phys. Rev. D **89**, 086010 (2014) [arXiv:1310.6334].